%% file: review-paper.tex
\documentclass[default,iicol]{sn-jnl}% Default with double column layout

\usepackage{graphicx}%
\usepackage{multirow}%
\usepackage{amsmath,amssymb,amsfonts}%
\usepackage{amsthm}%
\usepackage{mathrsfs}%
\usepackage[title]{appendix}%
\usepackage{xcolor}%
\usepackage{textcomp}%
\usepackage{manyfoot}%
\usepackage{booktabs}%
\usepackage{algorithm}%
\usepackage{algorithmicx}%
\usepackage{algpseudocode}%
\usepackage{listings}%
\usepackage[numbers]{natbib}
\usepackage{subcaption}
\usepackage{graphicx}
\usepackage{tikz}
\usepackage{pgfplots} 
\usetikzlibrary{patterns}
\theoremstyle{thmstyleone}%
\theoremstyle{thmstyletwo}%

\theoremstyle{thmstylethree}%

\begin{document}
	
	\title[Online Simulation at Machine Level: A Systematic Review]{Online Simulation at Machine Level: A Systematic Review}
	
	
	\author*[1,2]{\fnm{Darius} \sur{Deubert}}\email{darius.deubert@boschrexroth.de}
	
	\author[1]{\fnm{Lars} \sur{Klingel}}\email{lars.klingel@isw.uni-stuttgart.de}
	
	\author[2]{\fnm{Andreas} \sur{Selig}}\email{andreas.selig@boschrexroth.de}
	
	\affil[1]{\orgdiv{Institute for Control Engineering of Machine Tools and Manufacturing Units (ISW)}, \orgname{University of Stuttgart}, \orgaddress{\street{Seidenstraße 36},  \postcode{70174} \city{Stuttgart}, \country{Germany}}}
	
	\affil[2]{ \orgname{Bosch Rexroth AG}, \orgaddress{\street{Bgm.-Dr.-Nebel-Straße 2}, \postcode{97816} \city{Lohr am Main}, \country{Germany}}}


	\abstract{The importance of simulation at machine level in industrial environments is steadily increasing especially in the design and commissioning phase. Using models during the operation phase together with the real machine or plant is referred to as online simulation. Online simulation is used for system monitoring, predictive analyses, decision support or online optimization and therefore has various advantages and a wide field of applications. This paper has the aim to characterize online simulation at machine level in industrial automation focusing on key technologies and common applications. Therefore, a set of 60 relevant publications, which are focusing on this subject, is found by database search, expert consultation, and snowballing. As key technological aspects, the used model types, interfaces and platforms, and the aspects of initialization and synchronization are further investigated. The results are interpreted and limitations, knowledge gaps and future prospects are discussed. The potential of online simulation at machine level especially arises due to the increasing availability of component and machine models from the design and commissioning phase, which can be reused for online simulation. Remaining challenges are identified concerning implementation, simulation platforms, model maintenance and especially in the field of synchronization.}

	
	
	
	\maketitle
	
\section{Introduction}
\label{sec:intro}
\input{introduction.tex}

\section{Methods}
\label{sec:methods}
\input{methods.tex}

\section{Results}
\label{sec:results}
\input{results.tex}

\section{Discussion}
\label{sec:discussion}
\input{discussion.tex}

\section{Conclusion}
\label{sec:conclusion}
\input{conclusion.tex}

%\section*{Declarations}

%Some journals require declarations to be submitted in a standardized format. Please check the Instructions for Authors of the journal to which you are submitting to see if you need to complete this section. If yes, your manuscript must contain the following sections under the heading `Declarations':

%\begin{itemize}
%	\item Funding
%	\item Conflict of interest/Competing interests (check journal-specific guidelines for which heading to use)
%	\item Ethics approval 
%	\item Consent to participate
%	\item Consent for publication
%	\item Availability of data and materials
%	\item Code availability 
%	\item Authors' contributions
%\end{itemize}

%\noindent
%If any of the sections are not relevant to your manuscript, please include the heading and write `Not applicable' for that section. 

%%\bigskip
%%\begin{flushleft}%
%%	Editorial Policies for:
	
%%	\bigskip\noindent
%%	Springer journals and proceedings: %\url{https://www.springer.com/gp/editorial-policies}
	
%%	\bigskip\noindent
%%	Nature Portfolio journals: %\url{https://www.nature.com/nature-research/editorial-policies}
	
%%	\bigskip\noindent
%%	\textit{Scientific Reports}: %\url{https://www.nature.com/srep/journal-policies/editorial-policies}
	
%%	\bigskip\noindent
%%	BMC journals: %\url{https://www.biomedcentral.com/getpublished/editorial-policies}
%%\end{flushleft}



\bibliographystyle{sn-basic.bst} 
\bibliography{review-paper.bib}

\backmatter
\section*{Statements and Declarations}
\bmhead{Funding}
The work presented in this paper has been partly funded by the German Federal Ministry for Economic Affairs and Climate Action (BMWK) under the project 13IK001ZF “Software-Defined Manufacturing for the automotive and supplying industry \mbox{\url{https://www.sdm4fzi.de/}}”.

\end{document}

%% file: introduction.tex
Modern production plants are complex because of the increasing variety of products with various characteristics, while the number of human operators is decreasing.
To maintain today's production plants efficiently and to make optimal decisions, technical experts and autonomous intelligent systems need to have access to the exact plant state remotely.
For this, it is necessary to provide the internal state of the plant, its machines and its components as precisely as possible and to be able to predict the plant behavior in near future.
This problem is inter alia solved by online simulation at machine level. \cite{nakaya2013}

In the past, dynamic simulation based on first principle models (FPMs) were used offline in the system design phase, for testing and validation, and for process planning.
Due to advances in industrial communication technology, reduction of hardware costs and development towards flexible, automated and highly connected production environments, online simulation became more and more popular in the recent years. \cite{ruusu2017}

Online simulation means, that a process or system behavior is simulated during the system operation and the simulation is directly coupled with the real system to consider the same system state and real-time inputs that affect the real system \cite{davis1998}. 
Therefore, it is somehow connected to the real system, which could be a mechatronic component, a machine, a process or even a whole production plant or facility.

Online simulation has multiple applications and advantages \cite{nakaya2008, davis1998, pantelides2013, fagervik1988}:
\begin{itemize}
	\item Determination of quantities that cannot be obtained by measurement directly (virtual sensing)
	\item Substitution of expensive sensors
	\item Accessibility and easier visualization of system state and internal quantities
	\item Prediction of system behavior
	\item Validation of control laws
	\item Optimization of system operation
	\item Decision support
	\item Performance monitoring
	\item Diagnosis, fault detection and maintenance
\end{itemize}

In the past it was expensive and time consuming to build and maintain simulation models \cite{friman2012}.
Today, an increasing number of simulation models is existing from sizing tools, model based engineering and most recently also from virtual commissioning.
This development paves the way for online simulation, as the expenses for simulation models and platforms are decreasing and the possibilities of application in a modern \mbox{Industry 4.0} environment are extensive.

Early examples of online usage of simulation models are observers, like the Kalman Filter \cite{kalman1960}, or Model Predictive Control (MPC).
However, these applications mainly use simplified linear models, which are embedded deeply into mechatronic control systems.
Online simulation in industrial automation today comprises significantly more.
Applications reach from simple calculations for control tasks to complex simulations of entire manufacturing facilities.
As an online simulation is a digital replica of a machine or plant which is coupled to the real system in the operation phase, it is also referred to as simulation-based digital \mbox{twin \cite{martinez2018a}.}

However, there is no consistent understanding of online simulation in literature.
Different theoretical considerations and practical realization studies of online simulation have been published increasingly in the recent years, but there is no systematic overview of online simulation at machine level to the knowledge of the authors.

This article presents the state of the art on online simulation at machine level through a systematic literature review.
It focuses on the technical aspects, like initialization, synchronization, interfaces and platforms.
Furthermore, exemplary applications of online simulation are presented.
Through the reviewing process, knowledge gaps within the field of online simulation at machine level are identified, setting directions for future work.

The remainder of this article is structured as following:
In the next section, the used methods for the systematic review with a focus on literature research are described.
In section \ref{sec:results}, the results of the literature research are presented.
This section is structured based on different technical aspects.
Section \ref{sec:discussion} reviews the publications and summarizes limitations and challenges, as well as future potential.
In the end, a short conclusion is given.

%% file: methods.tex
This paper follows a systematic approach.
In particular, the methodology of this paper is based on Webster and Whatson \cite{webster2002}, Booth et al. \cite{papaioannou2016} and the PRISMA statement \cite{liberati2009}.

At first, a clear aim is defined.
The central research question is: 

\begin{quote}
	\normalsize
	\textit{Which approaches, methods and techniques concerning online simulation at machine level are existing and how are they applied in industrial practice?}
\end{quote}

This is further refined by the following questions, which should be answered by this article:

\begin{itemize}
	\item Which different methods, approaches and technologies are existing?
	\item Which different types of models are used for online simulation?
	\item How are the aspects of initialization and synchronization addressed in literature?
	\item Which platforms and interfaces are used for online simulation?
	\item Which applications exist for online simulation?
\end{itemize}

Furthermore, in the discussion section, remaining challenges and knowledge gaps are to be identified.

To find relevant publications, database search, snowballing (using the references between papers to identify additional papers) and expert consultation is used.
For the database search, the most important terms in the field of online simulation at machine level are identified.
This comprises different spellings of online simulation and tracking simulation, which is an important variation of online simulation, and terms to express the industrial context.
By connecting the terms with boolean operators the following advanced search string for database search results:

\begin{quote}
	 \texttt{( TITLE-ABS-KEY ( \{online simulation\} )  OR  TITLE-ABS-KEY ( \{on-line simulation\} )  OR  TITLE-ABS-KEY ( \{tracking simulation\} )  OR  TITLE-ABS-KEY ( \{tracking simulator\} ) )  AND  ( TITLE-ABS-KEY ( manufacturing )  OR  TITLE-ABS-KEY ( plant )  OR  TITLE-ABS-KEY ( "industrial automa*" ) ) } 
\end{quote}

Besides scopus, the databases IEEE Xplore and ACM Digital Library are chosen.
This search string yielded 191 results on scopus.
With slight variations concerning the syntax, the search string is applied to IEEE Xplore and ACM Digital Library and led to 117 results at IEEE Xplore and six results at ACM Digital Library.
The publications were collected in April 2023.

As the scope of this review is online simulation at machine level, especially in industrial automation, only publications, which focus on this subject should be included.
To narrow down the set of relevant publications to this field and filter irrelevant papers systematically, the following exclusion criteria are defined:
\begin{itemize}
	\item not referring to online simulation: publications, which do not refer to online simulation at all and therefore are completely irrelevant (affecting 57 publications)
	\item not in the context of industrial automation and manufacturing: publications referring to areas other than the mentioned ones, for example power plants or sewerage control (affecting 83 publications)
	\item not focusing on online simulation: publications, which emphasize on a subject, where online simulation is only handled as a side aspect and therefore not covered in detail (affecting 38 publications)
	\item Limitation to shop-floor level: publications, which are focusing too much on shop-floor level online simulation and where the content can hardly be transferred and applied at machine level (affecting 32 publications)
\end{itemize}
After removing duplicates, the articles were filtered by the application of these predefined exclusion criteria, resulting in a corpus of 42 papers.
The huge number of results, which do not refer to online simulation at all, are due to the ambiguous meaning and use of the terms \textit{online} and \textit{tracking} together with \textit{simulation}.
A major part of the research on online simulation before the 1990s was done at shop floor level, especially to support real-time scheduling \cite{krishnamurthi1993}.

A lot of these publications are irrelevant for online simulation at machine level and therefore excluded.
Most of the irrelevant papers were excluded during an initial abstract screening and the remaining papers were excluded after further analysis.

\begin{figure}
	\centering
	\includegraphics[width=0.9\linewidth]{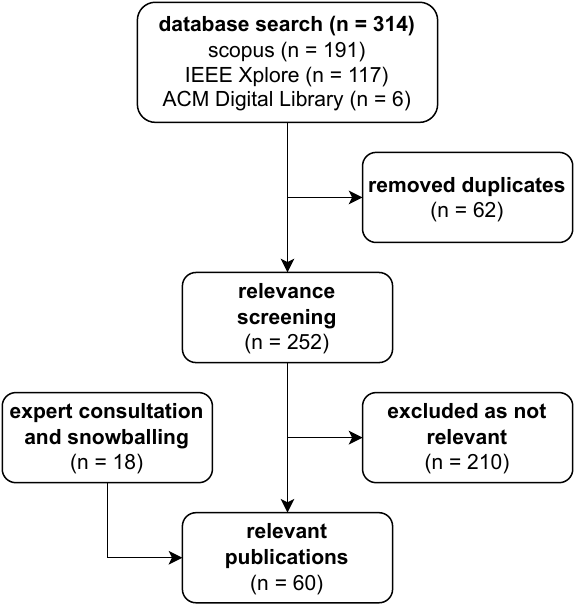}
	\caption{Accumulation of relevant papers (based on PRISMA flowchart \cite{liberati2009})}
	\label{fig:flowchart}
\end{figure}

By including nine articles, which were found by snowballing, and nine articles, which were found by expert consultation, a final set of 60 relevant publications is defined.
This step ensures that relevant papers, which are not found by the database search, are included in this review.
The overall process of literature selection is depicted in \mbox{Figure \ref{fig:flowchart}}.
The identified set of relevant publications was analyzed comprehensively and the results concerning the defined research questions are presented in the next section.

%% file: results.tex
Using simulation integrated and in parallel to the operation of a real system is called online simulation.
In the past, the main application at machine level was visualization and monitoring of process quantities, but as mentioned before, there is a wide variety of use-cases.
The main goal of online simulation at machine level is to support the optimal and undisturbed operation of the plant.

The following deals with different aspects of online simulation. 
It summarizes used model types, interfaces and platforms. 
Furthermore the aspects of initialization and synchronization are described and an overview of applications is given.

\subsection{Model types}
\input{models.tex}

\subsection{Initialization}
\input{initialization.tex}

\subsection{Synchronization and Tracking Simulation}
\label{sec:tracking-simulation}
\input{tracking-simulation.tex}

\subsection{Interfaces and platforms}
\label{sec:platforms}
\input{platforms.tex}

\subsection{Real-time behavior and temporal synchronization}
\input{real-time.tex}

\subsection{Application scenarios}
\label{sec:applications}
\input{application.tex}

%% file: models.tex
At first, different model types, which are used for online simulation, are compared.
Online simulation models should represent the real system as accurate as possible, but also be \mbox{re-configurable \cite{kain2009}}.
Modeling in general should need minimal effort, so there has to be found an optimum of accuracy and complexity.

\subsubsection{Discrete-event models}
In production and logistics, discrete-event simulation (DES) is used for analyzing, planning and optimizing processes.
DES models are based on instantaneous actions, which occur at a single moment and effect the system in its state. \cite{ferro2017}

DES is suitable for simulating flow of material or the capacity and utilization of production environments.
Therefore, DES models are often used to determine optimal production flow or waiting times on logistics or shop floor level.
DES is an effective tool for simulating alternative scenarios predictively \cite{zupan2021}.
Examples of DES models are Petri nets and final state machines \cite{jahn1996, bessey2004}.

On the other hand, continuous models are used for motions, speeds, cycle times and other processes, which often are real-time dependent and prevalent at machine level \cite{svensson2012}.
Nevertheless, to model the business and operation logic or communication state machines of automation components, DES models are also necessary at machine level.

\subsubsection{Hardware-in-the-Loop and Software-in-the-Loop models}
In the phase of plant engineering and commissioning, Hardware-in-the-Loop (HiL) and Software-in-the-Loop (SiL) models are used, for example for tests of control systems.
These models can also be used in the operation phase, for example as a reference for intended behavior.
As the HiL and SiL models are usually representing the operation logic of components or machines and only rough physical models, they are used in operation phase mainly for comparing the logical behavior of a system to its intended behavior \cite{kain2008}.
In combination with precise physical models, they can also be used for detailed monitoring and visualization of production processes and many other applications.

\begin{figure}
	\centering
	\includegraphics[width=0.75\linewidth]{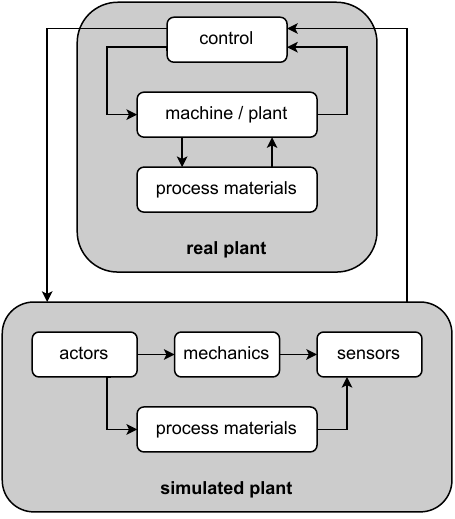}
	\caption{Structure of HiL online simulation (adapted from \cite{kain2008})}
	\label{fig:hil-structure}
\end{figure}
Kain et al. \cite{kain2008} present an exemplary structure of a HiL model for online plant simulation (see Figure \ref{fig:hil-structure}).
The simulation uses a model consisting of mechanics, sensors and actuators, as well as process materials.
Just like for HiL testing, the simulation environment is directly connected to the control, which could be a programmable logic controller (PLC) or a computer numerical control (CNC) and therefore needs corresponding interfaces \cite{kain2009}.

\subsubsection{Physical and statistical models}
Nakaya et al. \cite{nakaya2009} distinguish rigorous models, which are based on physical laws or phenomena, and statistical models, which are data-driven and consider the historical process data.
The rigorous models are commonly referred to as FPMs, which can be defined as models based on fundamental engineering, physics and chemistry principles \cite{pantelides2013}.

Statistical models are especially useful, if unknown phenomena, which could not be reconstructed by physical laws or are expensive to identify manually, have to be considered in the simulation model.
Examples for statistical models are artificial neural networks, look-up tables or generally different types of regression \cite{bergs2019}.
While FPMs can be reused easier, statistical models can be set up faster.
The disadvantages of FPMs are that the modeling and initial parameter tuning is complex and inexplicable phenomena cannot be considered.
Statistical models in contrast have to be maintained and provide less accurate estimations as long as they are in their learning phase \cite{nakaya2009}.
Additionally, they are not able to predict unusual behavior \cite{martinez2018a}.
When it comes to model adaption, data-driven statistical models are often newly learned, while physical models have to adjust their parameters \cite{martinez2017}.

Usual structures of statistical models are Multiple Linear Regression (MLR),
Principal Component Regression (PCR), Partial Least Squares (PLS) regression or Just-In-Time (JIT) \mbox{models \cite{nakabayashi2010}}.
The latter one is most suitable for applications with different operation ranges, as other methods tend to over-fitting.
Neural networks also suffer from over-fitting and furthermore are unintelligible in their meaning \cite{nakaya2011}.

When models are adjusted to a certain operation range to strong, they usually are worse applicable to other operation ranges.
JIT models store historical data in a database and only consider the relevant dataset for determining a local model.
The relevant dataset is chosen by a similarity factor, for example correlation, the Euclidean distance or other distance measures \cite{saptoro2014}.
By implementing a threshold, a new model is only built when the relevant dataset changes significantly.
JIT modelling is also called lazy learning or locally weighted learning.

\label{sec:hybrid-modeling}
\begin{figure*}
	\centering
	\hfill
	\begin{subfigure}{0.25\textwidth}
		\includegraphics[width=\textwidth]{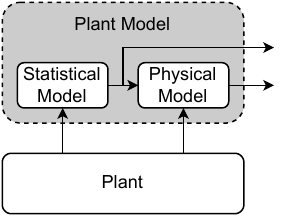}
		\caption{}
		\label{fig:combi1}
	\end{subfigure}
	\hfill
	\begin{subfigure}{0.25\textwidth}
		\includegraphics[width=\textwidth]{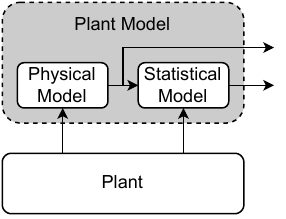}
		\caption{}
		\label{fig:combi2}
	\end{subfigure}
	\hfill
	\begin{subfigure}{0.25\textwidth}
		\includegraphics[width=\textwidth]{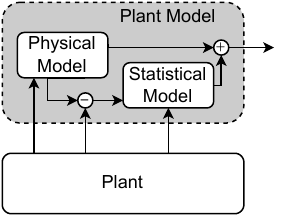}
		\caption{}
		\label{fig:combi3}
	\end{subfigure}
	\hfill
	\caption{Possible combinations of physical and statistical models in hybrid modeling (adapted from \cite{nakaya2013})}
	\label{fig:hybrid-combis}
\end{figure*}

To combine the advantages of both, first principle and statistical models, hybrid models are used. 
Generally, three types of combinations are possible (see Figure \ref{fig:hybrid-combis}):
\begin{itemize}
	\item statistical model as input for the physical model (Figure \ref{fig:combi1})
	\item statistical model as compensation for the physical model (Figure \ref{fig:combi2})
	\item statistical model processing the differences between physical model and real system  (\mbox{Figure \ref{fig:combi3}})
\end{itemize}

Nakabayashi et al. \cite{nakabayashi2010} applied a hybrid model consisting of a physical and a JIT statistical model on an online simulator for a steam reforming plant and proposed a new Mahalanobis distance-based JIT (MJIT). 
The Mahalanobis distance considers the covariance so it is correlation based, but does not need a hyper-parameter for weighting it relative to the distance measure.
Hence, only the remodeling threshold has to be set.
This is an advantage, as these hyper-parameters have to be set by trial-and-error, so they should be avoided or work as clear as possible \cite{saptoro2014}.

\subsubsection{Fluid dynamics modeling}
For fluid dynamics, which are necessary for industrial applications like bottling drinks or painting and casting steel, diverse methods are applicable.
Generally, models are divided into meshed and meshless discretization.
The most popular method using meshed discretization is Reynolds-averaged Navir Stokes, as it is strongly simplified.
For a higher resolution and turbulent flow, large-eddy-simulation or direct-numerical-simulations are common alternatives, but require a higher computational power.
Meshless methods can model turbulent flow more easily because of variable discretization.
As meshless methods are scalable, suitable for parallelization and have a high time step tolerance, they are more suitable for real-time simulation.
Examples are the smoothed particle hydrodynamics (SPH) method, the reproducing Kernel particle method (RKPM), and the finite point method (FPM).
\cite{krotil2016}

For online simulation, Krotil et al. \cite{krotil2016} define that before each iteration of model execution, the model is adjusted.
They distinguish fluid-specific model evolution, where sources and sinks are defined, and geometry-specific model evolution, which is derived from a CAD model.

\subsubsection{Automatically generated models}
Manual modeling in general is laborious and requires system experts.
Hence, efforts are made to generate models automatically.
As statistical models are generated by computation anyways, the main challenge of automatic online simulation model generation is the generation of underlying first principle models.
For this, information mapping algorithms are utilizing various information sources to map data into a simulation language.
Possible data and information sources are: \cite{martinez2018a}
\begin{itemize}
	\item Initial process design and equipment data sheets
	\item Piping and instrumentation diagram (P\&ID) in process industry
	\item Automation I/O tables
	\item 3D plant models
	\item Control application programs
\end{itemize}

As the real plant behavior may deviate from the automatically generated model, manual correction, applying an online parameter estimation or using tracking simulation, where the model parameters are optimized during operation, is recommended \cite{martinez2018a}.
Reasons for the deviation are that the system is built different from the design or due to degradation.
Santill\'an Martinez \mbox{et al. \cite{martinez2018a}} present a method, where they firstly generate a steady state model from preliminary process design data, which is then converted to a dynamic simulation model.
Alternatively, the dynamic simulation model can be built from the P\&ID or automation I/O tables and enhanced by further information.
Afterwards, the simulation models are connected to the manufacturing execution system (MES) and enterprise resource planning (ERP) systems for the operation phase. The method carries out the following steps:
\begin{enumerate}
	\item Automatic model generation
	\item Model integration within the plant
	\item Model optimization
	\item Online model parameter estimation
\end{enumerate}

Other terms, which are used in this context, are model adaption and model deployment.
In this context, Model adaption is the optimization of models, which were generated automatically or in another context, like in the design phase, for the use in online simulation.
The model integration within the plant is also referred to as model deployment.

Automatic model generation is still in its infancy, but is particularly important for economic efficiency and therefore for the widespread use in the production industry.

\subsubsection{Surrogate models}
As computing power is limited in mechatronic systems and online simulation at machine level requires real-time capability, high-fidelity simulation models are not suitable.
To address this, surrogate low-fidelity models are built.

To combine multiple models with different fidelity, multi-model-data-fusion is used \cite{bergs2019}.
By learning the fusion process with real measurements as ground truth, a hybrid of an existing model and a statistical model, which is represented by the fusion process, results.
Bergs and Heizmann \cite{bergs2019} present a generic approach for generating virtual sensors from high-fidelity models in that way.
In particular, low-fidelity models together with learning fusion operators are used to approximate the results of the high-fidelity simulation.
After a learning phase, the new combined model can be applied for online simulation, providing more accurate results than the low-fidelity model itself and being computationally cheaper than the high-fidelity model.

%% file: initialization.tex
One of the most important characteristics of online simulation is that it is initialized with the actual state of the system \cite{davis1998}.

To identify the actual state of a system, the following aspects have to be fulfilled \cite{fowler2004}:
\begin{itemize}
	\item Clear definition of the data forming the system state
	\item Availability of that data
	\item Sufficient quality of that data
	\item Sufficient update frequency of that data
\end{itemize}

This section summarizes different initialization strategies and presents a way of initialization data standardization.

\subsubsection{Initialization strategies} \label{sec:init-strat}
Online simulation has to be initialized with a state that is as close to the real system state as possible.
Hanisch et al. \cite{hanisch2005} present two possibilities for initialization.

The first option is to initialize an online simulation instance from a parent simulation instance, which is synchronized to the real system.
This requires an enhanced simulation platform, which is on the one hand able to copy a child simulation from a running simulation instance and on the other hand able to synchronize the parent simulation with the real system.
This option has the advantage that it is possible to gather the entire system state at one point of time.

The second option is to initialize simulation by using measurements of sensors or state information from system components.
This requires suitable connectivity and higher quality of data, but provides less restrictions regarding simulation platforms, as most of them are able to read and process initialization data \cite{hanisch2005}.
Especially if the measurement cycles are not synchronized, a higher measurement frequency enables a more accurate initialization.
However, in some systems a frequent cyclic measurement might not be possible, so this is not applicable in general.
Another common problem is that there might be relevant information or quantities, which are not captured by any sensor.
The determination of workpieces' location is often solved by retrofitting with radio-frequency identification (RFID) technology \cite{altaf2015}. 

Cardin et al. \cite{cardin2009} list another approach, which is a synthesis of the aforementioned options:
Observers use a model of the system to process the measurements and estimate the current state of the system, which is used for initialization.
Using observers is powerful, as it can be applied nearly everywhere and also provides the full state at any time. 
Cardin et al. \cite{cardin2012} emphasize the importance of centralizing plant data using an observer, which:
\begin{itemize}
	\item gathers the events occurring in the distributed manufacturing environment
	\item interpolates the behavior of the manufacturing units between the points of measurement
	\item makes the system state data available for the whole manufacturing environment
\end{itemize}

Another hybrid approach is presented by Santill\'an Martinez et al. \cite{martinez2015}.
They perform a rough database-based initialization with initial conditions, which is then tuned in a validation stage.
In the validation sequence, firstly the simulated state and then the model parameters are adjusted.
The offline tuning process can be repeated iteratively to get closer to the actual system state.
This process is called offline adaption and is followed by online adaption, which basically is a tracking simulation (see section \ref{sec:tracking-simulation}).
In this phase, the simulation is synchronized with the real plant.
If the discrepancy between simulation and real plant gets too strong during the online adaption, the offline adaption can be repeated.

Especially if models do not allow to directly set a certain state, a further possibility is to initialize the model with a standard state and use a simulated control system to align the simulated state to the actual state in an offline simulation \cite{martinez2018b}.

The choice of initialization strategy is depending on the availability and quality of initialization data \cite{hotz2006}.

\subsubsection{Initialization data}
As stated before, it is important to clearly define the system state.
There has to be a consistent understanding of which information and data is part of the system state and how the data has to be interpreted.

To enable interoperability between simulation systems there are various standards developed within the  Simulation Interoperability Standards Organization (SISO).
One of these standards is the core manufacturing simulation data information model (CMSD-IM), which focuses on interoperability between simulation and manufacturing applications.
The CMSD-IM provides information models and data structures in form of the Unified Modeling Language (UML) and the XML schema definition language (XSD).
Bergmann et al. \cite{bergmann2011} are mapping state information into CMSD-IM classes for the initialization of simulation models.
Specifically, the classes \textit{Resource, Part, Job, JobEffortDescription, Schedule, ScheduleItem} and \textit{ProcessPlan} are used.
Bergmann et al. \cite{bergmann2011} propose to extend the CMSD-IM by a timing property that indicates the process of a currently processed job.

Although the concepts refer to shop-floor level, they can be transferred to the machine level very well.
While jobs on shop-floor level are manufacturing or maintenance operations, on machine level they may be operation steps in a flexible production environment or steps/states within the execution of a specific program.
An exception are the interfaces, as Bergmann et al. assume to take the data from an Enterprise Resource Planning (ERP) system or a Manufacturing Execution System.
At machine level, such centralized systems often do not exist.
Nevertheless, the availability of real-time data is improved due to the industrial Internet-of-Things (IIoT), so that there are coming up many other possibilities for gathering the machine state.

%% file: tracking-simulation.tex
Synchronization is necessary to ensure that the online simulation can represent the current state of the real plant \cite{cardin2009}.
Due to wear, reconfiguration and malicious manipulation, the plant behavior is changing in its life cycle \cite{zipper2018}.
As manual synchronization would be costly and error prone, there have to be automatic synchronization mechanisms or at least monitoring systems, which detect deviations for maintenance personnel.

At machine level, this monitoring can be done utilizing the network.
Manufacturing machines typically consist of controllers, which are connected to actuators and sensors via fieldbuses.
Passive capturing only observes the present data stream in the network, while active capturing sends messages to gather information about the system state. \cite{zipper2018}

\begin{figure}
	\centering
	\hfill
	\begin{subfigure}{0.49\linewidth}
		\includegraphics[width=\textwidth]{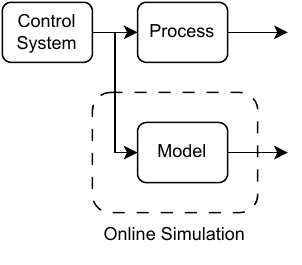}
		\caption{}
		\label{fig:online}
	\end{subfigure}
	\hfill
	\begin{subfigure}{0.49\linewidth}
		\includegraphics[width=\textwidth]{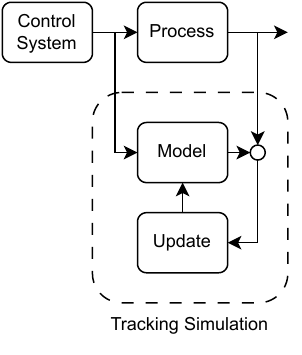}
		\caption{}
		\label{fig:tracking}
	\end{subfigure}
	\hfill
	\caption{Differences of standard online simulation (a) and tracking simulation (b) (adapted \mbox{from \cite{martinez2015}})}
	\label{fig:tracking-simulation}
	
\end{figure}
Besides monitoring,  adapting the simulation to be synchronous with the plant is a major challenge.
One way to solve the synchronization problem is by tracking simulation.

\subsubsection{Tracking simulation}
The concept of a tracking simulator was firstly presented by Nakaya et al. \cite{nakaya2006}.
They describe it as a dynamic process simulator, which works simultaneously with a target plant.
To ensure accuracy and minimize discrepancy between the real plant and the simulation, the model parameters are adjusted.
This is done by comparing sensor data with simulation results and updating the model parameters recursively, in a way that the simulation output converges to the real system output.
A visualization of this principle is pictured in \mbox{Figure \ref{fig:tracking-simulation}}. 
While basic online simulation does not use any feedback mechanisms (see \mbox{Figure \ref{fig:online}}), tracking simulation processes the difference between process output and simulated output to update the simulation model (see \mbox{Figure \ref{fig:tracking}}).

Reasons for diverging behavior between plant and simulation could be different operating conditions compared to the design base, changing characteristics due to degradation, or simply inaccurate simulation models \cite{nakabayashi2006}.

\begin{figure*}
	\centering
	\includegraphics[width=0.95\textwidth]{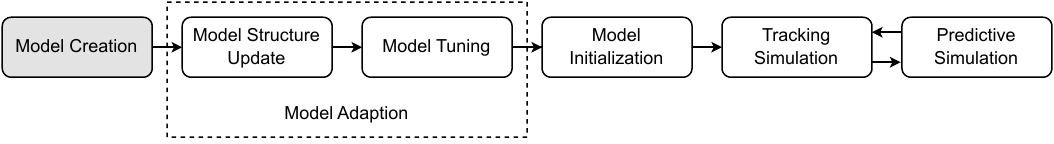}
	\caption{Stages of model adjustment for tracking simulation according to Santill\'an Martinez (adapted from \cite{martinez2017})}
	\label{fig:adaption-stages}
\end{figure*}
Santill\'an Martinez et al. \cite{martinez2015, martinez2017} distinguish different stages of model adjustment (see Figure \ref{fig:adaption-stages}).
In the model structure update, the model configuration is changed according to the plant, as the model created in the design phase could deviate.
This step often has to be carried out by experts.
In the model tuning stage, the model parameters are re-estimated.
At model initialization, the obtained system state and suitable parameters are transferred to the simulation.
In the case of tracking simulation, the state and the model parameters are continuously adjusted further on.
The tracking simulation instance may then be used as parent simulation for several predictive simulations.

Pantelides et al. \cite{pantelides2013} also propose three stages of model adaption:
\begin{enumerate}
	\item Offline model validation and parameter estimation
	\item Offline model-to-plant calibration
	\item Online calibration
\end{enumerate}

Nevertheless, initial conditions for tracking simulation must be chosen carefully, as the recursive determination of suitable parameters could take a long time, if the parameters are inaccurate in the beginning.
Hence, the initial values should be chosen as precise as possible. \cite{nakaya2007}

\subsubsection{Model parameter adjustment}
Besides common applications and advantages of online simulation, tracking simulation is useful to determine model parameters \cite{nakaya2008}.

The first approach was to adjust a parameter in a way inspired by a PI control.
For a parameter $\Lambda$, Nakaya et al. \cite{nakaya2006} determine the correction $\Delta \Lambda$ with:
\begin{equation}
	\Delta \Lambda = K_p e + K_i \int e\; dt \;,
\end{equation}
where $e$ is the difference between measurement and simulated output and $K_p$ and $K_i$ are feedback gains, which define how the feedback effects the parameter adjustment.

Friman et al. \cite{friman2012} also describe their tracking simulator as PI controller, whose error to minimize is the difference between simulated and real output and whose control value is an unknown parameter.

However, PI controllers can only be used, if the sign of the process gain is known.
Otherwise, wrong control gains can increase the simulation error.
Hence, the choice of control gains is not trivial.

Pietilä et al. \cite{pietila2013} are using a proportional and derivative part for an error function.
For the parameter update, they are using a predictor based estimation together with the error function.

An alternative solution is to use a sliding mode controller for the parameter adjustment.
This has the advantage that sliding mode control has a better disturbance rejection than PI-based controllers.
Furthermore, the sliding mode control needs less information about the system and is robust against uncertainties.
The disadvantages are that a sufficiently small step size and minimal delay is required.
As the sliding mode controller needs to know the derivatives of system and simulation outputs, an additional sliding mode observer is necessary in most cases. \cite{ruusu2017}

Another possibility is to record an output trajectory and solve an optimization problem to estimate optimal parameter adaptions, but this approach needs more computing power.
The advantage of an optimization algorithm is that it does not need that much tuning parameters \cite{pietila2013}.

For the model tuning, Santill\'an Martinez et al. \cite{martinez2017} utilize optimization algorithms that are applied on historical data.
In particular they present two variants, based on QNSTOP \cite{amos2020}, the Levenberg-Marquardt algorithm, and Broyden's method \cite{broyden1965}.
The Levenberg-Marquardt algorithm was also approved by Fagervik et al. \cite{fagervik1988}.
A problem of optimization algorithms for data reconciliation is that often analytic values for the Jacobian matrix of the system are necessary, which is not always given \cite{martinez2018b}.
Another problem is that scalar optimization algorithms may only find a local optimum.

In general, optimization problems can be solved by enumerative (brute-force), deterministic and stochastic approaches.
Enumerative ones are least efficient and only applicable to discrete or discretized problems, with a small solution space.
Deterministic approaches often need certain assumptions, so they are only applicable if these assumptions can be taken.
Härle et al. \cite{harle2021} therefore propose to use an evolutionary approach.
As optimization problems need a large computational effort, they are not suitable for real-time adaption.
Nevertheless, they can be used for periodical adaption of real-time online simulation.
\cite{harle2021}

\subsubsection{Enhancements of tracking simulation}
As the dynamic online adjustment of parameters often is limited, further mechanisms are necessary for the model adjustment.
The tracking simulator by Nakaya et al. \cite{nakaya2006, nakaya2007} was enhanced in \cite{seki2008}, where the simulation is divided into three parts:
\begin{itemize}
	\item The \textit{mirror model} is tracking the quantities of the real system by using measured values as explained before.
	\item The \textit{identification model} is adjusting certain model parameters periodically by data reconciliation using a least-squares method on a state-space representation. (also referred to as \textit{Dynamic Data Reconciliation} technique \cite{nakaya2011})
	\item An additional \textit{analysis model} is used for predictions, optimization, performance analysis and controllability improvement.
\end{itemize}
All parts rely on the same plant model.
The \textit{mirror model} and the \textit{identification model} are exchanging information in the form of variables and parameters \cite{nakaya2013}.

Another enhanced architecture is proposed by Santill\'an Martinez et al. \cite{martinez2018b}.
They also divide the simulation system into three simulators:
\begin{itemize}
	\item The \textit{online simulator} is running in parallel to the real plant.
	An initialization manager takes care of the model initialization.
	During the tracking simulation, the dynamic estimator adjusts the model parameters to align the simulation to the real process.
	The online simulator is also used for the initialization of child simulations as predictive or optimization simulators.
	\item The \textit{optimization simulator} runs offline model optimization on historical data.
	The optimized parameters are fed back to the online simulator.
	\item The \textit{predictive simulator} is used to be executed faster than real-time for forecasting of the process.
\end{itemize}

In addition to the process model, they use a control system model, as this enables predictive simulation and a control model is easier to integrate than a real control.
Hence, a replicate of the control application is used in the same simulation environment as the process model.

\subsubsection{State synchronization without model adaption}
\begin{figure}
	\centering
	\includegraphics[width=0.95\linewidth]{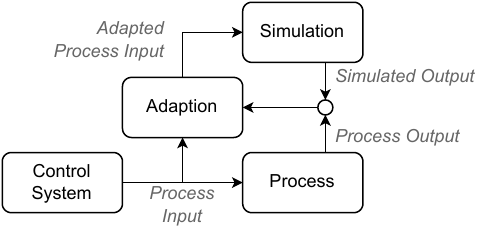}
	\caption{Synchronization approach by Zipper \cite{zipper2019}}
	\label{fig:zipper}
\end{figure}

Another approach, which also uses the difference between the simulation output and real output as feedback, but does not adapt the simulation model directly, is presented by Zipper \cite{zipper2019,zipper2021a,zipper2021b}.
The approach considers black-box Functional Mockup Interface (FMI) models, so the simulation system does not know about all model parameters and could not adapt them properly.
Instead, the system input is adapted for the simulation in a way that the difference between simulation output and real system output is compensated.
A visualization of this principle can be seen in Figure \ref{fig:zipper}.
Hence, this approach is closely related to tracking simulation.

For the minimization of the simulation discrepancy, an optimization algorithm over a certain time span is performed.
This time span is depending on the simulation steps needed for a changing input to affect the output.
This is depending on the number and constellation of simulation units in the co-simulation.
As optimization algorithms, Zipper validates L-BFGS and the derivative-free BOBYQA.

Alternatively to adapting the simulation input, the system's estimated state can be handled separately from the simulation result and therefore be updated by measurements.
This approach does also not require model adjustment but addresses state synchronization only indirectly.

%% file: platforms.tex
Predefined interfaces and existing software and hardware platforms are necessary for online simulation at machine level to establish in industrial practice.

\subsubsection{Hardware platforms}
Ebner et al. \cite{ebner2006} implemented a HiL test and simulation platform.
It comprises flexible hardware, such as a floating point digital signal processor (DSP), a field programmable gate array (FPGA), a pulse-width modulation (PWM) unit, an incremental encoder interface and a V/f-sensor interface, and high-speed communication interfaces.
To enable flexibility, expandability, and reusability, an appropriate operating system was developed. 
Even though the main application of the authors was to test control mechanisms for electric drives, the platform is perfectly suitable for online simulation, as models can be simulated in real-time and high-speed communication interfaces provide an appropriate connectivity towards the machine level and supervisory levels.
For modeling, they are using Dymola, which is a Modelica-based tool for modeling and simulation of mechanical, electrical, thermodynamic, hydraulic, pneumatic, thermal and control systems.
Dymola provides a code export functionality, which is allowing to convert the model into C-code for execution on the real-time platform.

Other applications do not utilize real-time targets for simulation.
Fagervik et al. \cite{fagervik1988} for example are using a personal computer, which is able to handle multiple input and output signals.
However, due to the used operating systems, simulations on a personal computer are not real-time capable in the most cases.

\subsubsection{Software platforms}
A general term that is used for simulation platforms in industry is computer-aided production engineering (CAPE).
CAPE tools are able to simulate manufacturing scenarios with various production resources and control functions \cite{svensson2012}.
The advantage of using CAPE tools for online simulation is that they provide numerous predefined models and, at best, a model of the plant is already present from the design or commissioning phase.
Furthermore, they already provide common interfaces.

Nakaya et al. \cite{nakaya2006,ishimaru2010} are using the dynamic simulation environment OmegaLand by the Japanese company Omega Simulation Co. \cite{seki2008}.

\begin{figure}
	\centering
	\includegraphics[width=0.9\linewidth]{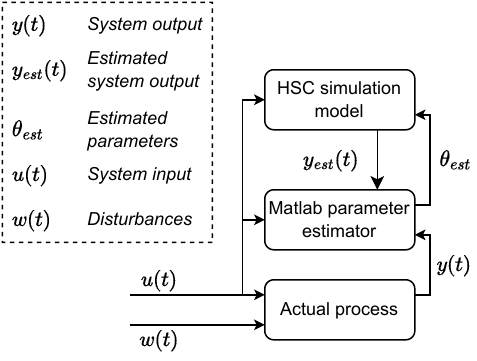}
	\caption{Chemical process simulation with continuous parameter adaption using Matlab (adapted from \cite{pietila2013})}
	\label{fig:pietila}
\end{figure}

Pietilä et al. \cite{pietila2013} are using the Outotec HSC simulator, which is a simulation tool for chemical processes.
They are using it as part of the tracking simulator and for evaluation.
The tracking simulator consists of a parameter estimator, which is implemented in Matlab, and the HSC simulation model.
An overview of the constellation is shown in Figure \ref{fig:pietila}.

\begin{figure}
	\centering
	\includegraphics[width=0.9\linewidth]{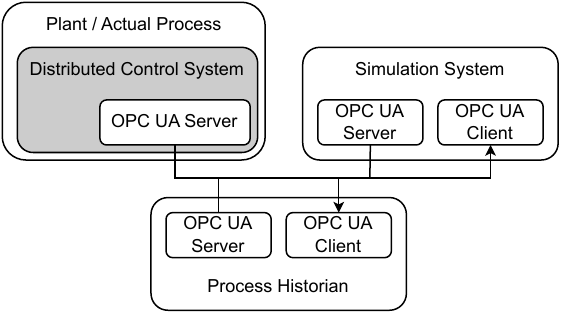}
	\caption{Online Simulation with OPC UA and process historian (adapted from \cite{martinez2017})}
	\label{fig:opc-ua}
\end{figure}

Martinez et al. \cite{martinez2015, martinez2017, martinez2018b} are using the process simulation Software Apros combined with OPC UA as an interface to the plant.
OPC UA is a client-server-based communication standard, which provides a unified interface for heterogeneous data sources in Industry 4.0 environments.
The communication structure for tracking simulation with OPC UA is depicted in Figure \ref{fig:opc-ua}.
The OPC UA historical access is used for the model tuning before model initialization.
As in many cases the time-steps are irregular, the data series has to be interpolated.
This is a disadvantage for simulations with fixed step sizes and optimization algorithms relying on evenly distributed time series.

Petschnigg et al. \cite{petschnigg2018} realized an interface to robot operating system (ROS).
As a simulation platform, they use CoppeliaSim (formerly V-REP).
With an extending script, path planning for robotic applications is possible.
It is connected to CoppeliaSim with the CoppeliaSim client API.
The API client can control the simulation state, retrieve and edit simulation data, like object positions, and enqueue planning targets, which are target poses for the tool center point.

In industrial automation, Modelica models are popular.
Modelica is an object-oriented simulation language, which is used by multiple simulation platforms and hence, provides many ready-to-use models \cite{ebner2006}.
Further popular simulation platforms in industrial automation are ISG-virtuos, machineering iPhysics, Matlab/Simulink and Siemens tools like Siemens plant simulation.
Some of them come with real-time simulation targets, which can be used for HiL simulation.

To sum up, domain specific simulation platforms were often used and adapted to the requirements of online simulation.
A software platform especially intended for online simulation does not exist to the knowledge of the authors.
However, many existing simulation platforms already provide suitable interfaces for online simulation.
In future, standards like the FMI will play an important role, as they enable the co-simulation of various existing component models that cant set up a machine or plant model.

Some of the reviewed implementations also use programming languages like C/C++, Java and Python directly.
In practice, newly coding component or machine models with general purpose programming languages is expensive and therefore not suitable for the most applications, but a way to be more flexible in research and testing.

\subsubsection{Cloud services}
Depending on the model complexity and the optimization mechanisms, online simulation might be computationally costly.
In this case, cloud computing services are considered, as they allow scalability and parallelization.
Especially for small and medium-sized companies, third party cloud services like Amazon Web Services (AWS) and Microsoft Azure are attractive, as they provide the infrastructure and are paid per use. \cite{hofmann2022}

\begin{figure*}
	\centering
	\includegraphics[width=\textwidth]{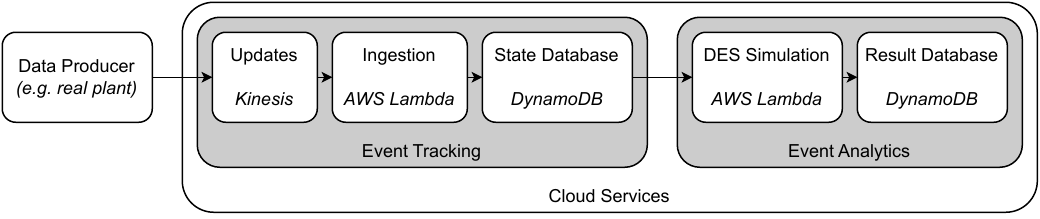}
	\caption{Exemplary deployment of DES in AWS cloud environment (adapted from \cite{hofmann2022})}
	\label{fig:aws}
\end{figure*}
Hofmann et al. \cite{hofmann2022} present an exemplary model for the deployment of a DES simulation on the AWS cloud (see Figure \ref{fig:aws}).
The data producer could be a real production system, which collects information with its sensors and triggers corresponding events.
These events are captured by AWS Kinesis, which handles the data stream and acts as a communication interface.
The events are processed first-in-first-out by AWS Lambda.
AWS Lambda updates a Dynamo DB database, which stores the events.
The DES model is also implemented as AWS Lambda function.
According to the occurring events, the state variables are changed in the Dynamo DB database.
The AWS Lambda functions are executed depending on cell updates in the Dynamo DB.
The simulation time is synchronized with the real system.

\subsubsection{Interfaces}
The synchronization of online simulation to a real system, as well as the cooperation with other mechatronic or virtual systems require consistent interfaces.
While on shop-floor level human-machine-interfaces to the simulation are used not only for supervision and monitoring but also for manually providing system inputs \cite{kadar2010}, the online simulation at machine level has to focus especially on machine-to-machine communication.

Machine-to-machine communication is divided into different architectures, such as service-oriented architectures (SOA), Representational State Transfer (REST) and message-oriented architectures \cite{muller2021}.
In practice, the communication is realized by standardized \mbox{Internet-of-Things (IoT)} protocols on the one hand and APIs to proprietary apps and platforms on the other hand \cite{lim2022}.

Apart from this, it is also possible to connect the simulation to the mechatronic system using fieldbuses and Industrial Ethernet protocols.
Industrial fieldbuses are necessary, if the application requires real-time capability.
\mbox{Zipper et al. \cite{zipper2018}} for example use the Ethernet-based PROFINET and map the process data to the corresponding simulation inputs.
Test access points, which are located between the controller and the considered components, capture the data from the communication network directly.
This has the advantage that the high amount of data is processed close to the process and does not have to be spread through the factory network or even further \cite{zipper2018}.

Cardin et al. \cite{cardin2011} implemented a simulation system containing a discrete event simulation as an observer and initialization basis for predictive online simulation instances.
An OPC server is used to synchronize the event-based observer simulation to the real system.
The OPC server sends events, when predefined variables are changing.
For the initialization of the predictive online simulation instances a database containing production data is used.

Martinez et al. \cite{martinez2018b} propose an architecture, where three different simulators (an online simulator, an optimization simulator, and a predictive simulation) are connected to a historical data repository, which stores process and simulation data in a database.
For the historical data repository, they use a Prosys OPC UA historian, which also uses an SQL-based database.

OPC UA has a service-oriented architecture.
With OPC UA Pub/Sub, a cyber physical system can register to a topic as publisher or subscriber and a broker coordinates the data distribution.

An alternative to OPC UA is Message Queuing Telemetry Transport (MQTT), which is a message-oriented protocol that offers multiple advantages for asynchronous data streams.
An MQTT broker running on a server manages the message distribution inside a network.
MQTT is supported by several open-source libraries. \cite{muller2021}

In the case of Müller et al. \cite{muller2021}, different online simulation instances for different manufacturing units exchange their state information via MQTT.
They used an update interval of 200 ms, which is acceptable for the communication with MES systems or non-real-time components, but cannot compete with protocols used in mechatronic systems internally, like industrial Ethernet.

%% file: real-time.tex
Online simulation at machine level is subjected to real-time constraints, as the simulation results are needed at a certain point of time at the plant and may be irrelevant afterwards \cite{davis1998}.
Real-time behavior of a system refers to the ability to keep deadlines and, especially regarding simulation, to be able to synchronize to the real clock.
Both aspects play a vital role within online simulation.

Nakaya et al. \cite{nakaya2013} see the application of their tracking simulator for a petrochemical plant on minute-level to hour-level and even in the static domain, which corresponds to day-level.
That means, that their proposed tracking simulator is not suitable for real-time use on machine level, where control cycles are in the order of milliseconds.

HiL models, which usually are connected to a control system via a fieldbus, underlie hard real-time requirements.
They often operate on the control's cycle, which is in the order of milliseconds for PLCs.

Also for real-time decision making, it is important to provide information by time.
In this case, the time scale is depending on the application, but as several possibilities have to be simulated until certain decision points, the simulation has to be performed much faster than real time.
There has to be found an optimum of simulation time scale and simulation speed concerning simulation accuracy, which is related to a longer simulation duration, and expressiveness, which is shrinking with the simulation's duration due to drifting of the real system's behavior \cite{bessey2003a}. 
The simulation time scale and simulation speed can be influenced by computing power, simulation step size and level of detail. 

\subsubsection{Parallel and predictive online simulation}
Online simulation can be run in parallel to the real system or sped up to predict the system behavior.
Important applications of parallel simulation are monitoring, diagnosis and robust operation.
The diagnosis aspect of parallel online simulation is important for the service and maintenance of a plant, as deviations to the intended behavior can be detected and tracked more easily.
By automatically replying to these deviations, a robust operation is provided \cite{kain2008}.

Predictive online simulation has to run faster than real time and, in contrast to parallel online simulation, also model and simulate the control, as future inputs and reference values have to be considered.
However, a control model is often also used for parallel online simulation, as it is easier to integrate the control into the simulation in that way \cite{martinez2018b}.
Predictive online simulation aims to forecast failures in the plant behavior and also determines degradation, which enables early scheduling of service tasks.
Another application of predictive online simulation is process optimization.
Different control options can be evaluated by simulation and the best option is chosen for the process.
In that way, decision support is given and optimal process control is enabled. \cite{kain2008}

Cardin et al. \cite{cardin2009} use predictive online simulation for decision making in holonic manufacturing systems (HMS).
They presented a novel approach of proactive production activity control systems using online simulation \cite{cardin2011}.
Instead of periodically launching predictive simulations, like for model predictive control and other proactive production activity control systems (see \mbox{Pujo et al. \cite{pujo2004}}), the predictive simulation can be launched on demand.
By simulating different options of control, the predictive simulation is used to support decision making.
For accurate initialization of the predictive online simulation, a real-time simulator is used as an observer.
This approach is used often, as it provides a straightforward and accurate way to obtain the system state for the initialization of further simulation instances (see section \ref{sec:init-strat}).

To reduce computing time for predictive simulation, Manivannan and Banks \cite{manivannan1991} propose a combination with a knowledge base, which processes simulation results and observations of the real plant to assist a discrete event simulation.

\subsubsection{Temporal Synchronization}
Cardin et al. \cite{cardin2011} present a simple mechanism to synchronize an event driven simulation to the real system.
When an event occurs in the simulation, it waits for the real event to occur before the simulation is continued.
When an event occurs in the real system and the simulation is not waiting for it, the simulation is continued, but the next occurrence of the event in the simulation is suppressed.

Manivannan and Banks \cite{manivannan1991} synchronize their discrete event simulation by synchronizing the event list.

Bessey \cite{bessey2003b} proposes a method, which adapts the length of successive predictive simulations to cause similar simulation horizons.
This is necessary for the simulation results to be better comparable.

%% file: application.tex
In this section, some of the most important applications for online simulation at machine level and relevant applications of the aforementioned techniques are presented, including implementation details.

Pantelides et al. \cite{pantelides2013} distinguish between closed-loop and open-loop online model-based applications (OMBAs).
Closed-loop control OMBAs provide direct feedback to the physical system for process control.
These could be virtual sensors or model predictive controllers.
On the other hand, open-loop advisory OMBAs have no direct feedback and are used mainly as support systems, such as monitoring, error detection or \mbox{forecasting \cite{martinez2018b}}.

Another distinction is made between OMBAs for plant monitoring, which extend the information of the plant provided by the sensors, and OMBAs for plant forecasting.

The simulation can run continuously (e.g. for process monitoring), on demand (e.g. for decision making), or at scheduled times (e.g. for performance monitoring). \cite{pantelides2013}

\subsubsection{Virtual assembly supported system}

\begin{figure}
	\centering
	\includegraphics[width=0.9\linewidth]{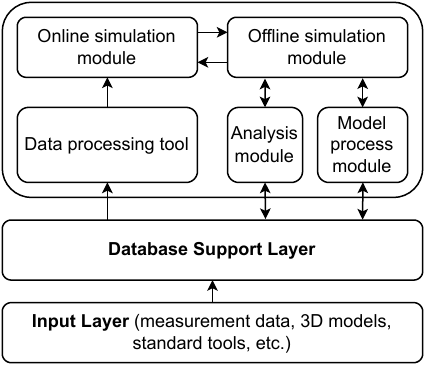}
	\caption{Structure of the symbiotic simulation proposed by Meng et al. \cite{meng2013}}
	\label{fig:vass}
\end{figure}

A virtual assembly supported system for gas turbine manufacturing is presented by Meng et al. \cite{meng2013}.
The system consists of an input layer, including measurement data, 3D models, etc., a database support layer, and a layer consisting of an online simulation module, a data processing module, an offline simulation module, an analysis module and a model process module (see Figure \ref{fig:vass}).

The online simulation is responsible for real-time guidance and supervision of the assembly process.
This means, it finds deviations and errors, generates warnings and provides a feedback mechanism for the offline simulation.
It is implemented as a real-time HiL simulation.

The offline simulation deals with process planning and analysis and is implemented in the virtual space.
Both modules are cooperating, as the offline simulation is a preview and reference for the online simulation.

The simulations use actual data from the assembly system and enhance the assembly by feeding back process information.
There is diverse cooperation between the modules, where information is exchanged and resources are shared.
Such compositions are known as symbiotic \mbox{simulations \cite{scheer2021}}.

\subsubsection{Decision making}
Online simulation has been used for decision making at MES level for a long time \cite{iassinovski2008}.
The simulation is used to evaluate different scenarios and therefore simplify the decision making.

Cardin et al. \cite{cardin2011} apply their proactive activity control, which basically is an online simulation system for decision making, to a holonic manufacturing system consisting of different working stations and a transportation system.
Different orders are created and handled by the manufacturing system.
When a new order is created, a specific number of transporters have to be assigned.
By predictive simulation of the manufacturing system, the choice of transporters was optimized in a way that the productivity was increased by 30\% in average.

Yoshitani et al. \cite{yoshitani1991} apply this principle to a reheating furnace for slab heating.
They are using a high-fidelity model for predictions within a long range and a low-fidelity model fort short term predictions.
The approach is comparable to model predictive control, where an optimal future control trajectory is simulated and the determined control input is applied to the plant.

Other examples of online simulation for decision support are Manivannan and Banks \cite{manivannan1991}, Bessey \cite{bessey2004} and Krishnamurthi and Vasudevan \cite{krishnamurthi1993}.

Besides operational decision making, where online simulation mainly is used to evaluate short-term decisions by predictive simulations, decision support is also needed for engineering services.
This comprises condition monitoring and asset life-cycle management.
Tjahjono et al. \cite{tjahjono2013} present an online simulation framework, which consists of a reliability estimator, an engineering service simulation model and a condition monitoring system.
The reliability estimator uses the condition monitoring data to determine a likelihood of failure. 
This indicator is used to plan maintenance actions.
The engineering service model enables predictive simulation to achieve an optimal maintenance activity.
A data capturing module collects data from the real system to synchronize to the actual state. This state is used to initialize predictive online simulation.
Their experiments showed that online simulation compared to traditional methods allows an improved monitoring, especially concerning systems with dynamic behavior and perturbation.
This is a valuable advantage for flexible manufacturing systems in the context of Industry 4.0.

However, online simulation for decision making is less prevalent at machine level, except model-based control methods.
The major reason might be that machines had a fixed repetitive operation in the past, where usually no adaptions were made.
This will change within Industry 4.0 due to individualization and lot size one approaches.
Additionally, adaptions need experts at machine level, while at shop floor level simple heuristics and adequate information are usually sufficient.
Hence, decisions and optimizations at machine level might be more complicated than shop floor level decision support and optimized production scheduling.

\subsubsection{Simulation-based early warning system}
Hotz et al. \cite{hotz2006} use online simulation for an early warning system.
The idea of simulation-based early warning systems is, to collect data from production control centers, according databases or even from CNCs or PLCs and use them together with simulation models to provide information about the future behavior of a production system.
An early warning system is designed independently from the simulation tool and the simulation has to be embedded invisibly, so that the user of the early warning system does not need to think about the simulation.
Hence, parameterization, initialization and the execution of the simulation have to be automated.

\subsubsection{Tracking simulator for steam reforming plant}
In the process industry, the system behavior is changing over long time due to degradation of catalyst activity and process drift \cite{nakaya2013}.
The idea of a tracking simulator came up thinking about transferring offline simulators from the process \mbox{industry (petrochemical,} steel or paper plants etc.) into an adapting online simulation \cite{nakaya2006}.
That is why Nakaya et al. \cite{nakaya2006, nakabayashi2006, nakabayashi2010, nakaya2007, nakaya2008, nakaya2009, nakaya2011, nakaya2013} are using a steam reforming plant for a fuel cell as a demonstrator, where it was commercially implemented later on.
For this, they use statistical models and FPMs for a hybrid model simulation \cite{nakaya2009}.
The statistical model is fed with measured values and data from the simulation.

Ishimaru et al. \cite{ishimaru2010} also applied the tracking simulator to a depropanizer process with a distillation column.

The main purpose of the tracking simulator in this case is to serve as a virtual sensor for different chemico-physical quantities.
Besides, it provides possibilities for trend forecasts and optimized operation conditions.
Nakaya et al. \cite{nakaya2013} propose the use of tracking simulation together with model predictive control (MPC) and real-time optimization (RTO) as MPC and RTO need current estimations for unstable parameters.

\subsubsection{Vibration Reduction}
Due to energetic efficiency and material costs, lightweight machines are built more often.
However, lightweight machines are prone to vibration, which has to be avoided.
Measuring resonant frequencies of a machine before the commissioning depending on machine positions is costly.
Sekler et al. \cite{sekler2009, sekler2012} propose a method to determine resonant frequencies from a simulation model, whose behavior is calculated continuously on the control device during the operation.
In particular, predefined models are coupled based on the machine position to determine the resonant frequencies for the current machine position.

\subsubsection{Wood-frame panel prefabrication}
Altaf et al. \cite{altaf2015} implemented an online simulation for a wood-frame panel prefabrication facility, which consists of an RFID data acquisition module, a central database and a discrete event simulation.
Here, the simulation is applied to a production line, consisting of multiple machines, but the principle can also be applied at machine level, for example for a machine tool.

The database contains the building panel information from 3D models, the production schedule, and the panel locations, which are fed from the data acquisition module.

The online simulation is implemented in \mbox{Simphony.NET}, where the production steps are modeled as tasks, which are done by machines with workers as resources.
The simulation initializes with data from the database and runs based on task-time formulas, which are developed empirically and determine the processing time based on the panels geometry.
Data, which could not be obtained automatically, like the number of human workers, are put in via a user interface.

The simulation is used to predict the performance of the production line and to evaluate the real performance.
This is necessary, as the panel size is an insufficient indicator for the production time \cite{altaf2015}.
Transferring this to machine level, online simulation provides new measures, which can be used to determine the performance of a machine.
These new measures are perfectly suitable for pay-per-use approaches.

\subsubsection{Digital Twin for CNC machine tool}
CNC machine tools are prevalent in manufacturing.
Luo et al. \cite{luo2019} present a digital twin for a CNC machine tool, which provides precise simulation, self-sensing, self-adjustment, self-prediction, and self-assessment.
The mentioned aspects play a key role in Industry 4.0 concerning the concepts of cyber physical production systems and decentralized intelligence.
One important pillar therefor is modeling and simulation.
In particular, \mbox{Luo et al. \cite{luo2019}} use a system level simulation consisting of a descriptive model and an algorithm model, which accesses real-time data for fault prediction using a fuzzy neural network.

\subsubsection{Flexible robot manufacturing}
Traditionally, robot movements are typically hard-coded and sensors detecting the workpieces and obstacles are not available.
However, a lot of tools exist, which enable the simulation of robots, for example to test applications and code.
Some tools even support automatic path generation.
Virtual robots are used to reduce the companies' time-to-market and to improve the production \mbox{processes. \cite{petschnigg2018}}

Petschnigg et al. \cite{petschnigg2018} mention several difficulties of using simulation for robot operation:
\begin{itemize}
	\item Dynamically changing environment impedes collision detection, when paths are planned by the simulation. 
	Hence, either some kind of feedback information of the environment via sensors or a dedicated cell, where the environment is fixed, is necessary.
	\item Dynamic change of the workpiece during the simulated process is hard to include into 3D simulation with CAD models.
	\item The necessary degree of detail is crucial, but hard to obtain.
	Therefore, often it is tried to get an as detailed and as accurate model as possible.
	\item  The dynamic behavior of the end effector is hard to pre-calculate.
\end{itemize}

In their work, they connect a production environment with a parametrizable simulation, which can calculate robot trajectories adapted to the current state of the environment.
The size and pose of obstacles and workpieces and the current joint configuration of the robot are fed into the simulation system, where they are used for path planning.

\subsubsection{Online collision avoidance}
Flexible production systems demand new safeguarding solutions, especially when online adaptions and reconfiguration to the production system increase the risk of a collision.
In \cite{schumann2013} and \cite{hoher2013} a concept for an online collision avoidance system is presented.
In such a system an online simulation is used, to calculate the motion of the real production systems in parallel.
In the virtual world the parts are surrounded by bounding volumes.
Typically those bounding volumes are larger than their real world counterpart to detect collisions virtually before real collisions occur.
Alternatively, the online simulation is run predictively.
If a collision is detected in the online simulation, this information can be used to stop the production system or initiate another reaction to prevent the collision.
In \cite{klingel2023} such a solution is realized for a robot cell.
This makes the close and flexible cooperation of three industrial robots possible.

\subsubsection{Operation monitoring}
In flexible manufacturing systems, a plant often consists of different production units, which dynamically execute jobs on work pieces.
For dynamic collaboration, the controllers need detailed online state information of the machines and robots.
For this, Jahn \cite{jahn1996} uses synchronized online simulation to reflect the state and behavior of production cells.
By this, simple sensors can be used to update and synchronize the online simulation with respect to the mechatronic system and detailed state information is provided at any time by the online simulation.
In particular, Jahn \cite{jahn1996} applies DES, but generally, any type of simulation may be useful to determine system information.
The model type has to be chosen depending on the kind of information that is necessary for operation monitoring.

%% file: discussion.tex
In this section, the investigated publications on online simulation at machine level are interpreted and the results are discussed.
This includes knowledge gaps and limitations as well as future potential.

\subsection{Current limitations and remaining challenges}
Fowler et al. \cite{fowler2004} identified two major challenges for online simulation in 2005.
Firstly, synchronized factory models with real-time data access should be run permanently.
This challenge is weakened in Industry 4.0 due to high availability of accurate real-time data and increasing computational power on edge and cloud devices, but still existent.
Furthermore, there still is a lack of synchronization techniques.

Secondly, the factory models have to be built automatically.
The excessive costs of model development is one major reason, why online simulation is not applied in practice \cite{pantelides2013}.
Santill\'an Martinez et al. \cite{martinez2018b} also state that a focus for future work should be set in reducing the efforts for modeling and integration.
This challenge is still present, but there already are multiple activities in solving this problem \cite{bergmann2020, dammasch2010}.

Besides automatic model generation, automatic validation of simulation models is a current challenge that is necessary for online simulation on the one hand and could be solved by online simulation on the other hand, also concerning models that are not only used during the operation phase.

In the following, further limitations and challenges are derived.

\subsubsection{Implementation}
A lot of theoretical concepts have been presented, which can be used or transferred to machine level.
However, there is still a need for practical concepts, to implement online simulation in industrial practice \cite{martinez2018b}.
Especially at machine level, the implementation is difficult, as real-time process data has to be handled.
If the simulates takes place in the cloud, suitable interfaces have to be established in the mechatronic system.
If the simulation is run on field level, a device has to be found that:
\begin{itemize}
	\item has enough computing power for the simulation
	\item can access the real system's inputs
	\item has interfaces to provide the generated information to the control system and to analysis and maintenance tools
	\item provides an environment, where the simulation is conducted, administered and maintained
\end{itemize}
Online simulation needs to access the real system's inputs in any case.
These inputs comprise endogenous inputs, which are inputs coming from the control system, and exogenous inputs, which are mostly unwanted impacts, which the system has no control of \cite{davis1998}.
The exogenous inputs are hard to consider, as the mechatronic systems often do not provide sensors for these quantities.

Implementation challenges also include real-time capability.
Especially cloud solutions and high-fidelity models have to be improved regarding this to meet the future requirements.

In total, there are no general and clear approaches on how to proceed to realize an appropriate online simulation for a specific application or problem, but only separate theoretical approaches with individual implementation examples.

\subsubsection{Platforms}
Simulation platforms are helping to overcome implementation challenges, as they provide appropriate simulation and modeling tools, various interfaces to the mechatronic system and sometimes also real-time capable simulation devices (see section \ref{sec:platforms}).

ISG-virtuos for example is real-time capable and can be  connected to common fieldbus systems, like CANopen, Ethernet IP, EtherCAT, Profinet, Profibus, or Powerlink.
Furthermore, it supports the use of FMI models and has a C\texttt{++} SDK.
To be integrated directly into the mechatronic system and to separate simulation from modeling, ISG provides the Realtime Target, which is a compact, cabinet-based industrial PC.
Hence, ISG-virtuos as an example already provides the most important features for online simulation.

However, to the knowledge of the authors there is no simulation platform that is explicitly dedicated for online simulation.
As described, current simulation platforms already provide important features for online simulation, but especially methods for synchronization and instantiation of child models, with the ability to run in different simulation speeds, have to be developed.

\subsubsection{Real-time capability}
Real-time capability is a vital aspect for online simulation at machine level.
Using existing platforms saves implementation efforts, but may restrict the simulation system concerning real-time capability depending on the real-time behavior of the simulation platform.
Besides interfaces, where industrial Ethernet is recommended for real-time communication, the operating system and the computing power of the simulation device are the main bottlenecks.
Many operating systems cannot provide the necessary real-time capability at all.
In other cases the models are to complex to be executed with real-time conditions.
Depending on the simulation platform, there even is an overhead additionally to the model complexity, for example due to a Simulation Master for co-simulation or for administrative functionality.
To provide cheap computing power to handle complex models, cloud services are used.
This solves the problem of computing power, but enhances the problem of real-time capable communication.

For these reasons, many of today's implementations of online simulation at machine level are not real-time capable.
In the considered set of 60 relevant publications, only 16 publications address real-time capability and nine publications consider HiL models, which usually are real-time capable (see Figure \ref{fig:bar-chart}).

An example, where real-time capable simulation of mechatronic systems is already realized, is virtual commissioning with HiL models.
This example could serve as a role model for online simulation at machine level and simulation platforms can be reused for both applications.

Furthermore, approaches are necessary to reduce the model complexity and to save computing power, maintaining the informative value of the online simulation. 

\subsubsection{Model maintenance}
As the modeled plants and machines are changed and reconfigured over time, also the models for online simulation have to be adapted.
This exceeds the synchronization techniques mentioned in section \ref{sec:tracking-simulation}, as models not only are optimized towards gradual changes in behavior but may have to be completely reconfigured.
As online simulation aims to reduce maintenance efforts of production plants, the maintenance effort for the simulation itself has to be minimized.
Today, maintenance of models is complicated and entails a lot of effort, as large parts are done manually.
This problem not only exists in the context of online simulation but generally in the field of digital twins, where the concept of an evolving digital twin \cite{lee2023,lin2021,edington2023} came up recently.
Companies are having high expenses for keeping their digital twins up to date.
This is a huge challenge to be solved in future.

Model sustainability is highly related to this and entails solvability, maintainability and tractability \cite{pantelides2013}.
Hence, models should not be too complex but also not too simple.
Furthermore, they should be adaptable and modular by \mbox{design \cite{lugaresi2018}.}
The distinction between model maintenance, model adaption and model synchronization is vague in literature.

Another aspect affecting model maintenance is model reuse.
For this, models should be flexible in their application, so that they can not only be used for online simulation, but also for virtual commissioning or in the design phase.
As the models might have to be adapted for the different applications, version management is necessary.

\subsubsection{Synchronization}
\begin{figure}
	\centering
	\begin{tikzpicture}
	\begin{axis}[
	ybar,
	ymin=0,
	x=7ex,
	bar width=3ex,
	ymajorgrids,
	xtick pos=left,
	ytick pos=left,
	symbolic x coords = {real-time capability, HiL models, model reuse, state synchronization, model synchronization},
	nodes near coords,
	x tick label style = {rotate=60, anchor=east},
	]
	\addplot[fill=gray!80,draw=black!80] coordinates { (real-time capability,16) (HiL models,9)
		(model reuse,17)  (state synchronization,42) (model synchronization, 26) };
	\end{axis}
	\end{tikzpicture}
	\caption{Number of the relevant publications found (see methods, section \ref{sec:methods}) addressing an issue (total number of considered publications: 60)}
	\label{fig:bar-chart}
\end{figure}
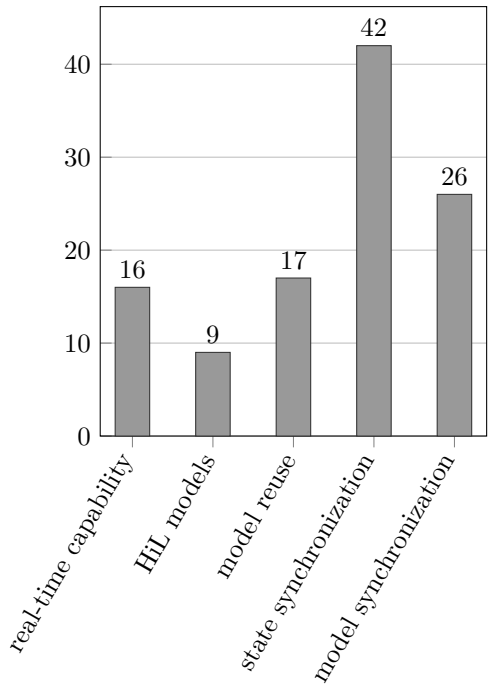

The aspect of synchronizing the online simulation to the real system is essential.
State synchronization, where only the state variables are adjusted to fit the real system state, is distinguished from model synchronization, where the simulation model is adapted corresponding to the actual system behavior.
Figure \ref{fig:bar-chart} shows that 42 of the 60 investigated publications address the aspect of state synchronization and another 26 publications also address model synchronization.
The shown statistics are not quantitatively representative but show a clear qualitative direction.

Model synchronization is especially necessary, if the behavior of a system is changing over time.
As mentioned before, this could be due to degradation, changing environment or changing operation conditions.
Additionally, inaccurate models are improved by model synchronization.
This is why the concept of tracking simulation is a fundamental step towards the usage of online simulation in industrial automation.

The state synchronization method by Zipper has several disadvantages compared to regular tracking simulation, as the models themselves are not tuned and the simulated state only is corrected indirectly.
If model parameters are incorrect due to bad modeling or changing behavior, the synchronization method by Zipper compensates this by changing the simulation input in a way that may be unrealistic, which could cause the internal state to be unrealistic.
However, the implementations of tracking simulation presented in the literature are not suitable for co-simulation with black-box models.
As the modeling effort is reduced if the models of machines are made up from component models and composed in a co-simulation environment, and component models provided by the component manufacturers are supposed to be more accurate, black-box model co-simulation could supposably be the future standard for machine modeling in machinery manufacturing.
The approach by Zipper is intended especially for this scenario.

A limitation of the current synchronization techniques is that they do not consider the uncertainty of measurement, which is used for the model and state correction.
Also, if there is high noise, the state and model are adjusted too often and too strong. 
Using the Kalman filter as a role model, the uncertainty of measurement and the uncertainty of simulation should be considered to achieve statistically optimal state and model adjustments.
Furthermore, determining the uncertainty of the state estimation would help for further processing.
Apart from this, Bessey \cite{bessey2003b} proposes a method, which weights simulation results of predictive simulations, which were conducted successively, as the different simulations used different real-time data for initialization.

An explicit gap exists in the combination of model synchronization with HiL models, which is not addressed by the reviewed literature at all.
Model synchronization and real-time capable simulation is addressed only by Santill\'an Martinez et al. \cite{martinez2018b} and Bessey \cite{bessey2003a}, which only deals with the need for model synchronization and real-time capability independently, but not in combination.
At machine level, real-time capability is especially important, since the simulation results should be used in the control system directly.
HiL models play a key role here, as they already exist from testing and virtual commissioning and are real-time capable innately.
Feedback from the real system and proper reaction in form of synchronization is necessary to ensure the accuracy and reliability of online simulation.

Besides state and model synchronization, temporal synchronization is a challenge.
The synchronization of different clocks itself has many solutions, but the synchronization of multiple interacting simulation environments for example, is an upcoming problem that is to be \mbox{solved \cite{hofmann2022}}.
Applying distributed simulation speeds up the simulation, but orchestrating multiple local simulations is complex \cite{bessey2003a}.

\subsubsection{Processing of simulation results}
As online simulation at machine level is expanding, it enables new possibilities due to the supply with new types of data and information at field and control level.
Hence, questions on the use of this data and information are coming up, like how to interpret the simulation data and how to use these results.
The results could be of quantitative manner, for example giving estimations for physical quantities, or qualitative manner, for example indicating if certain system states are avoided.
Especially the latter is not obvious, but important for retrospective analysis and safety critical systems.
The future online simulation systems have to provide appropriate interoperability, to enable such analyses \cite{bessey2003a}.

In general, retrospective analysis, which focuses on monitoring of the mechatronic system, is distinguished from prospective analysis, which focuses on the future operation, for example by validating or comparing control laws, estimating future performance or determining optimal maintenance activities.
Online simulation supports the use of data analysis technologies, since it provides more accurate and diverse data in the role of a digital twin.

\subsection{Trends and future potential}
\begin{figure*}
	\centering
	\pgfkeys{/pgf/number format/.cd,1000 sep={}}
	\begin{tikzpicture}
	\begin{axis}[
	ybar,
	ymin=0,
	x=2.8ex,
	bar width=1.5ex,
	ymajorgrids,
	xtick pos=left,
	ytick pos=left,
	xmin = 1987.1,
	xmax = 2022.9,
	xtick distance = {1},
	ytick distance = {1},
	x tick label style = {rotate=90, anchor=east},
	]
	\addplot+[ybar,fill=gray!80,draw=black!80] coordinates {
		(1988,1) 
		(1989,0) 
		(1990,0) 
		(1991,2) 
		(1992,0) 
		(1993,1) 
		(1994,0) 
		(1995,0) 
		(1996,1) 
		(1997,0) 
		(1998,1) 
		(1999,0) 
		(2000,0) 
		(2001,0) 
		(2002,0) 
		(2003,2) 
		(2004,2) 
		(2005,1) 
		(2006,4) 
		(2007,1) 
		(2008,4) 
		(2009,4)
		(2010,3)  
		(2011,3)  
		(2012,4)  
		(2013,5)  
		(2014,0)  
		(2015,1)  
		(2016,2)  
		(2017,3)  
		(2018,6)  
		(2019,2)
		(2020,0)  
		(2021,5)    
		(2022,2)  
	};
	\end{axis}
	\end{tikzpicture}
	\caption{Number of publications per year, which are considered as relevant}
	\label{fig:bar-chart2}
\end{figure*}
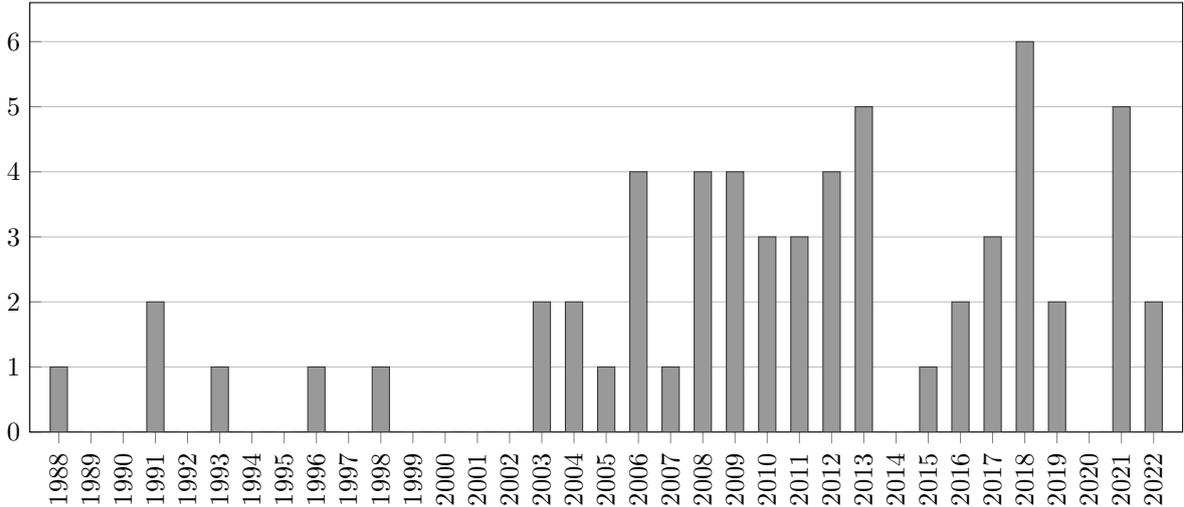

As seen in Figure \ref{fig:bar-chart2}, the number of relevant publications on online simulation at machine level is increasing, although there is high variation from year to year.
The potential of online simulation at machine level especially arises, as an increasing number of existing models reduces the obstacle of modeling expenses to use online simulation.
Also the number and extent of suitable simulation platforms is increasing.

Additionally, online simulation is a form of a digital twin.
The use of digital twins is proclaimed as a major pillar for digitalization and is seen as one of the most important technology trends of the recent years.
Hence, also the importance of online simulation is rising.

Common applications of online simulation at machine level are presented in section \ref{sec:applications}.
Important applications of the past were decision making support and virtual sensing, while decision making support takes place rather at shop-floor level.
Nevertheless, machine level simulation is also necessary for shop-floor level decision making.
Virtual sensing is part of decision making support, but mainly creates new possibilities concerning system control and enables to substitute physical sensors, where they are expensive or hard to integrate.

Another application of online simulation is model validation \cite{lugaresi2018}.
This is done by applying the model for online simulation and comparing the simulation results with the actual system behavior.
By this, manual effort for model validation is reduced.

\subsubsection{Model reuse}
Due to the increasing presence of various models from the design, engineering and commissioning phase, further utilization of these existing models reduces the expenses for the application of first principle models.
Many models are not used after the design phase.
Online simulation supports the concept of efficient re-utilization and consistent use of digital twin components across the whole life cycle, as models from the design phase are reused.
As seen in Figure \ref{fig:bar-chart}, 17 out of the 60 considered publications address model reuse.

Model reuse exceeds the limits of machines or plants, as component models or submodels in general can also be used across different machines or plants.
This further improves economic efficiency and creates new value chains and business \mbox{models. \cite{rosen2020}}

Model reuse requires fixed interfaces and encapsulation of models.
Current efforts are also aimed on easy integration into the plant.
All applications have to be considered right from the beginning of the development and the mentioned aspects have to be implemented by design.

Summed up, models should be reused consistently and efficiently across machines or plants and life-cycle phases.
Reusing models for online simulation is a crucial step towards this goal.
On machine level, especially HiL and SiL models from virtual commissioning could be reused.

\subsubsection{Online simulation as form of a digital twin}
As mentioned before, online simulation of a system is a form of the system's digital twin.
Generally, a digital twin is a virtual representation of an asset.
This digital representation ranges from ERP relevant data representations to 3D CAD models.
At machine level, online simulation is an important pillar of corresponding digital twins.
Online simulation includes models of the machine, aggregates data and information concerning the machine, and even generates further data and information by simulation.

According to Kritzinger \cite{kritzinger2018}, a digital twin must have bidirectional automatic data flow between the real asset and the virtual counterpart.
If there is no automatic data flow, the virtual representation is just a digital model.
Transferring this to simulation, offline simulation only belongs to digital models, but using simulation of a plant online and feeding data back to the real plant can be referred to as a digital twin.
This integration is covered in section \ref{sec:platforms}.

As a lot of data from mechatronic systems is collected in the context of digital twins, online simulation plays a vital role as an easily accessible data source.
For accessibility, often cloud environments, SQL databases, and IoT protocols are used (see \cite{zupan2021,lim2022, hofmann2022}).
Alternatively an OPC UA server could be integrated \mbox{(see \cite{martinez2017, martinez2018b, martinez2015,svensson2012, luo2019, ruusu2017})}, which seems to be the future de facto standard in industrial communication.

Additionally to simply collect data, online simulators with model synchronization can also detect changes in the behavior.
Also the diversity of collected data is enhanced.
Together with the ability of predictive simulation, these aspects of online simulation are supporting service and maintenance.

By enabling deeper insights into the machine's or plant's state and behavior remotely, teleoperation is supported.
As the number of human operators in production should be reduced for more efficient manufacturing and experts are necessary to operate the complex plants, teleoperation is becoming an increasingly vital factor. 

\subsubsection{Comparison to alternatives}
One major difference of online simulation based on FPMs compared to learning statistical models, like neural networks, is that online simulators based on FPMs have plausible internal behavior.
This is a big advantage, as diverse phenomena could be interpreted more clearly and conclusions on the causes can be drawn.
The results of model updates can be interpreted physically and therefore be used for service and maintenance.
Furthermore, behavior, which is not known from the past, can be predicted by the first principle models, as they are not based on historical data.
However, it is more complicated to guarantee suitable model adaption for FPMs.

Another alternative for monitoring purposes are observers or Kalman filters.
Kalman filters should be treated as role models for online simulation regarding the synchronization, as Kalman filters are processing the real system output in a statistically optimal way to adapt the estimated system state.
Also model synchronization can be realized by Kalman filters, by including model parameters into the state space.
However, they are limited to linearized state space models and machine models in industrial automation could not always be easily transferred to state space models.

%% file: conclusion.tex
This work is intended as an overview of the state-of-the-art of online simulation at machine level but also as an indication for remaining problems and future work.

Online simulation at machine level provides multiple possibilities.
As shown, there are many applications for online simulation and its relevance is increasing, due to Industry 4.0 and the prevalence of digital twins.
Furthermore, the obstacle of expensive modeling is reduced by automatic model generation and reuse of existing models from previous life-cycle phases.
Due to these reasons, online simulation has an immense potential for future applications at machine level.

As described in section \ref{sec:discussion}, there are still many challenges that have to be overcome to use online simulation efficiently.
One major aspect is synchronization of the simulation to the actual plant or machine.
Another challenge is the implementation.
Multiple theoretical considerations concerning initialization, synchronization, suitable platforms and real-time behavior have been summarized in this article, but there are many practical problems left, which have to be solved by future work.

%% file: review-paper.bbl
\begin{thebibliography}{79}
\providecommand{\natexlab}[1]{#1}
\providecommand{\url}[1]{{#1}}
\providecommand{\urlprefix}{URL }
\providecommand{\doi}[1]{\url{https://doi.org/#1}}
\providecommand{\eprint}[2][]{\url{#2}}
 \bibcommenthead

\bibitem[{Altaf et~al(2015)Altaf, Liu, Al-Hussein, and Yu}]{altaf2015}
Altaf MS, Liu H, Al-Hussein M, et~al (2015) {Online simulation modeling of
  prefabricated wall panel production using RFID system}. In: {2015 Winter
  Simulation Conference (WSC)}, IEEE, pp 3379--3390

\bibitem[{Amos et~al(2020)Amos, Easterling, Watson, Thacker, Castle, and
  Trosset}]{amos2020}
Amos BD, Easterling DR, Watson LT, et~al (2020) {Algorithm 1007: QNSTOP —
  quasi-Newton algorithm for stochastic optimization}. ACM Transactions on
  Mathematical Software (TOMS) 46(2):1--20

\bibitem[{Bergmann and Stra{\ss}burger(2020)}]{bergmann2020}
Bergmann S, Stra{\ss}burger S (2020) {Automatische fg Modellgenerierung --
  Stand, Klassifizierung und ein Anwendungsbeispiel}. Ablaufsimulation in der
  Automobilindustrie pp 333--347

\bibitem[{Bergmann et~al(2011)Bergmann, Stelzer, and
  Stra{\ss}burger}]{bergmann2011}
Bergmann S, Stelzer S, Stra{\ss}burger S (2011) {Initialization of simulation
  models using CMSD}. In: {Proceedings of the 2011 Winter Simulation Conference
  (WSC)}, IEEE, pp 2223--2234

\bibitem[{Bergs and Heizmann(2019)}]{bergs2019}
Bergs C, Heizmann M (2019) {Kombination unterschiedlicher
  Modellierungsans{\"a}tze f{\"u}r die betriebsbegleitende Simulation
  industrieller Prozesse}. at-Automatisierungstechnik 67(3):183--192

\bibitem[{Bessey(2003{\natexlab{a}})}]{bessey2003a}
Bessey T (2003{\natexlab{a}}) Needs and proposals for theoretical research on
  on-line simulation. In: Summer Computer Simulation Conference, Society for
  Computer Simulation International; 1998, pp 459--466

\bibitem[{Bessey(2003{\natexlab{b}})}]{bessey2003b}
Bessey T (2003{\natexlab{b}}) On-line simulation: Towards new statistical
  approaches. In: Summer Computer Simulation Conference, Society for Computer
  Simulation International; 1998, pp 453--458

\bibitem[{Bessey(2004)}]{bessey2004}
Bessey T (2004) {Implementation of on-line simulation with the colored Petri
  net simulator RENEW}. In: {2004 IEEE International Conference on Systems, Man
  and Cybernetics (IEEE Cat. No. 04CH37583)}, IEEE, pp 5019--5024

\bibitem[{Broyden(1965)}]{broyden1965}
Broyden CG (1965) {A class of methods for solving nonlinear simultaneous
  equations}. Mathematics of computation 19(92):577--593

\bibitem[{Cardin and Castagna(2009)}]{cardin2009}
Cardin O, Castagna P (2009) {Using online simulation in Holonic manufacturing
  systems}. Engineering Applications of Artificial Intelligence
  22(7):1025--1033

\bibitem[{Cardin and Castagna(2011)}]{cardin2011}
Cardin O, Castagna P (2011) {Proactive production activity control by online
  simulation}. International Journal of Simulation and Process Modelling
  6(3):177--186

\bibitem[{Cardin and Castagna(2012)}]{cardin2012}
Cardin O, Castagna P (2012) {Myopia of service oriented manufacturing systems:
  benefits of data centralization with a discrete-event observer}. Service
  Orientation in Holonic and Multi-Agent Manufacturing Control pp 197--210

\bibitem[{Dammasch et~al(2010)Dammasch, Kaupp, and Rabuser}]{dammasch2010}
Dammasch K, Kaupp H, Rabuser M (2010) {Eine Automatische Modellgenerierung zur
  simulationsgest{\"u}tzten Planung und Optimierung von robotergesteuerten
  Fertigungsprozessen}. Integrationsaspekte der Simulation: Technik,
  Organisation und Personal KIT Scientific Publishing, Karlsruhe pp 53--60

\bibitem[{Davis(1998)}]{davis1998}
Davis WJ (1998) {On-line simulation: Need and evolving research requirements}.
  Handbook of simulation 465:516

\bibitem[{Ebner et~al(2006)Ebner, Ganchev, Gragger, and Pirker}]{ebner2006}
Ebner A, Ganchev M, Gragger JV, et~al (2006) {Real Time Platform for Rapid
  Prototyping and On-line Simulation of Digital Controllers for Electrical
  Drives}. SAE Transactions pp 120--125

\bibitem[{Edington et~al(2023)Edington, Dervilis, {Ben Abdessalem}, and
  Wagg}]{edington2023}
Edington L, Dervilis N, {Ben Abdessalem} A, et~al (2023) {A time-evolving
  digital twin tool for engineering dynamics applications}. Mechanical Systems
  and Signal Processing 188:109971. \doi{10.1016/j.ymssp.2022.109971}

\bibitem[{Fagervik et~al(1988)Fagervik, Konstari, and von
  Schalien}]{fagervik1988}
Fagervik K, Konstari O, von Schalien R (1988) {Control of batch evaporative
  crystallization of sugar by means of adaptive simulation}. In: {1988 American
  Control Conference}, IEEE, pp 677--683

\bibitem[{Ferro et~al(2017)Ferro, Ord{\'o}{\~n}ez, and Anholon}]{ferro2017}
Ferro R, Ord{\'o}{\~n}ez REC, Anholon R (2017) {Analysis of the integration
  between operations management manufacturing tools with discrete event
  simulation}. Production Engineering 11:467--476

\bibitem[{Fowler and Rose(2004)}]{fowler2004}
Fowler JW, Rose O (2004) {Grand challenges in modeling and simulation of
  complex manufacturing systems}. Simulation 80(9):469--476

\bibitem[{Friman and Airikka(2012)}]{friman2012}
Friman M, Airikka P (2012) {Tracking simulation based on PI controllers and
  autotuning}. IFAC Proceedings Volumes 45(3):548--553

\bibitem[{Hanisch et~al(2005)Hanisch, Tolujew, and Schulze}]{hanisch2005}
Hanisch A, Tolujew J, Schulze T (2005) {Initialization of online simulation
  models}. In: {Proceedings of the Winter Simulation Conference, 2005.}, IEEE,
  pp 9--pp

\bibitem[{H{\"a}rle et~al(2021)H{\"a}rle, Barth, and Fay}]{harle2021}
H{\"a}rle C, Barth M, Fay A (2021) {Operation-parallel adaptation of a
  co-simulation for discrete manufacturing plants}. In: {2021 26th IEEE
  International Conference on Emerging Technologies and Factory Automation
  (ETFA)}, IEEE, pp 1--8

\bibitem[{Hofmann et~al(2022)Hofmann, Lang, Reichardt, and
  Reggelin}]{hofmann2022}
Hofmann W, Lang S, Reichardt P, et~al (2022) {A brief introduction to deploy
  Amazon Web Services for online discrete-event simulation}. Procedia Computer
  Science 200:386--393

\bibitem[{Hoher et~al(2013)Hoher, Neher, and Verl}]{hoher2013}
Hoher S, Neher P, Verl A (2013) {Collision: Impossible - Echtzeitfähige
  3D-Kollisionskontrolle bei mehrkanaliger Bearbeitung}. SPS IPC Drives 2013

\bibitem[{Hotz et~al(2006)Hotz, Hanisch, and Schulze}]{hotz2006}
Hotz I, Hanisch A, Schulze T (2006) {Simulation-based early warning systems as
  a practical approach for the automotive industry}. In: {Proceedings of the
  2006 Winter Simulation Conference}, IEEE, pp 1962--1970

\bibitem[{Iassinovski et~al(2008)Iassinovski, Artiba, and
  Fagnart}]{iassinovski2008}
Iassinovski S, Artiba A, Fagnart C (2008) {SD Builder{\textregistered}: A
  production rules-based tool for on-line simulation, decision making and
  discrete process control}. Engineering Applications of Artificial
  Intelligence 21(3):406--418

\bibitem[{Ishimaru et~al(2010)Ishimaru, Nakaya, and Ohtani}]{ishimaru2010}
Ishimaru S, Nakaya M, Ohtani T (2010) {An application of tracking simulator to
  depropanizer process}. In: {Proceedings of SICE Annual Conference 2010},
  IEEE, pp 1486--1489

\bibitem[{Jahn(1996)}]{jahn1996}
Jahn G (1996) {Modeling concepts for data reduction in control of manufacturing
  systems}. Cybernetics \& Systems 27(3):223--234

\bibitem[{K{\'a}d{\'a}r et~al(2010)K{\'a}d{\'a}r, Lengyel, Monostori,
  Suginishi, Pfeiffer, and Nonaka}]{kadar2010}
K{\'a}d{\'a}r B, Lengyel A, Monostori L, et~al (2010) {Enhanced control of
  complex production structures by tight coupling of the digital and the
  physical worlds}. CIRP annals 59(1):437--440

\bibitem[{Kain et~al(2008)Kain, Heuschmann, and Schiller}]{kain2008}
Kain S, Heuschmann C, Schiller F (2008) {Von der virtuellen Inbetriebnahme zur
  Betriebsparallelen Simulation}. atp edition 50(08):48--52

\bibitem[{Kain et~al(2009)Kain, Dominka, Merz, and Schiller}]{kain2009}
Kain S, Dominka S, Merz M, et~al (2009) {Reuse of HiL simulation models in the
  operation phase of production plants}. In: {2009 IEEE International
  Conference on Industrial Technology}, IEEE, pp 1--6

\bibitem[{Kalman(1960)}]{kalman1960}
Kalman RE (1960) A new approach to linear filtering and prediction problems.
  Transactions of the ASME--Journal of Basic Engineering 82(Series D):35--45

\bibitem[{Klingel and Verl(2023)}]{klingel2023}
Klingel L, Verl A (2023) {Simulationsbasierte Online-Absicherung von
  CNC-gesteuerten Industrierobotern}. Fortschritt-Berichte VDI pp 74--82

\bibitem[{Krishnamurthi and Vasudevan(1993)}]{krishnamurthi1993}
Krishnamurthi M, Vasudevan S (1993) {Domain-based on-line simulation for
  real-time decision support}. In: {Proceedings of the 25th conference on
  Winter simulation}, pp 1304--1312

\bibitem[{Kritzinger et~al(2018)Kritzinger, Karner, Traar, Henjes, and
  Sihn}]{kritzinger2018}
Kritzinger W, Karner M, Traar G, et~al (2018) {Digital Twin in manufacturing: A
  categorical literature review and classification}. Ifac-PapersOnline
  51(11):1016--1022

\bibitem[{Krotil et~al(2016)Krotil, Richter, and Reinhart}]{krotil2016}
Krotil S, Richter C, Reinhart G (2016) {Online-simulation of fluidic processes
  in early design of plant development using SPH}. CIRP Annals 65(1):161--164

\bibitem[{Lee et~al(2023)Lee, Kim, and Yoon}]{lee2023}
Lee Y, Kim S, Yoon K (2023) {Class Abstraction and Upcasting for Self-evolving
  Digital Twin System}. In: {2023 International Conference on Electronics,
  Information, and Communication (ICEIC)}, pp 1--3,
  \doi{10.1109/ICEIC57457.2023.10049945}

\bibitem[{Liberati et~al(2009)Liberati, Altman, Tetzlaff, Mulrow, G{\o}tzsche,
  Ioannidis, Clarke, Devereaux, Kleijnen, and Moher}]{liberati2009}
Liberati A, Altman DG, Tetzlaff J, et~al (2009) {The PRISMA statement for
  reporting systematic reviews and meta-analyses of studies that evaluate
  health care interventions: explanation and elaboration}. Annals of internal
  medicine 151(4):W--65

\bibitem[{Lim et~al(2022)Lim, Lee, Yoo, and Yoon}]{lim2022}
Lim Y, Lee YK, Yoo J, et~al (2022) {An Open Source-based Digital Twin Broker
  Interface for Interaction between Real and Virtual Assets}. In: {2022 13th
  International Conference on Information and Communication Technology
  Convergence (ICTC)}, IEEE, pp 1657--1659

\bibitem[{Lin et~al(2021)Lin, Jia, Yang, Xiao, Lan, Shi, Zeng, and
  Li}]{lin2021}
Lin TY, Jia Z, Yang C, et~al (2021) {Evolutionary digital twin: A new approach
  for intelligent industrial product development}. Advanced Engineering
  Informatics 47:101209

\bibitem[{Lugaresi and Matta(2018)}]{lugaresi2018}
Lugaresi G, Matta A (2018) {Real-time simulation in manufacturing systems:
  Challenges and research directions}. In: {2018 Winter Simulation Conference
  (WSC)}, IEEE, pp 3319--3330

\bibitem[{Luo et~al(2019)Luo, Hu, Zhang, and Wei}]{luo2019}
Luo W, Hu T, Zhang C, et~al (2019) {Digital twin for CNC machine tool: modeling
  and using strategy}. Journal of Ambient Intelligence and Humanized Computing
  10:1129--1140

\bibitem[{Manivannan and Banks(1991)}]{manivannan1991}
Manivannan S, Banks J (1991) {Real-time control of a manufacturing cell using
  knowledge-based simulation}. In: {1991 Winter Simulation Conference
  Proceedings.}, pp 251--260, \doi{10.1109/WSC.1991.185622}

\bibitem[{Meng et~al(2013)Meng, Zhang, and Wang}]{meng2013}
Meng X, Zhang L, Wang M (2013) {Symbiotic simulation of assembly quality
  control in large gas turbine manufacturing}. In: {AsiaSim 2013: 13th
  International Conference on Systems Simulation, Singapore, November 6-8,
  2013. Proceedings 13}, Springer, pp 298--309

\bibitem[{M{\"u}ller et~al(2021)M{\"u}ller, Mielke, Pavlovskyi, Pape, Masik,
  Reggelin, and H{\"a}berer}]{muller2021}
M{\"u}ller M, Mielke J, Pavlovskyi Y, et~al (2021) {Real-time combination of
  material flow simulation, digital twins of manufacturing cells, an AGV and a
  mixed-reality application}. Procedia CIRP 104:1607--1612

\bibitem[{Nakabayashi et~al(2006)Nakabayashi, Fukano, Onoe, Nakaya, and
  Ohtani}]{nakabayashi2006}
Nakabayashi A, Fukano G, Onoe Y, et~al (2006) {Application of tracking
  simulator to reforming process}. In: {2006 SICE-ICASE International Joint
  Conference}, IEEE, pp 1871--1875

\bibitem[{Nakabayashi et~al(2010)Nakabayashi, Nakaya, Ohtani, Chen, Wang, and
  Li}]{nakabayashi2010}
Nakabayashi A, Nakaya M, Ohtani T, et~al (2010) {A process simulator based on
  hybrid model of physical model and Just-In-Time model}. In: {Proceedings of
  SICE Annual Conference 2010}, IEEE, pp 1497--1501

\bibitem[{Nakaya and li(2013)}]{nakaya2013}
Nakaya M, li X (2013) {On-line tracking simulator with a hybrid of physical and
  Just-In-Time models}. Journal of Process Control 23:171–178.
  \doi{10.1016/j.jprocont.2012.06.007}

\bibitem[{Nakaya et~al(2006)Nakaya, Fukano, Onoe, and Ohtani}]{nakaya2006}
Nakaya M, Fukano G, Onoe Y, et~al (2006) On-line simulator for plant operation.
  In: Proceedings of the World Congress on Intelligent Control and Automation
  (WCICA), pp 7882 -- 7885, \doi{10.1109/WCICA.2006.1713505}

\bibitem[{Nakaya et~al(2007)Nakaya, Kawaguchi, Onoe, Watanabe, and
  Ootani}]{nakaya2007}
Nakaya M, Kawaguchi K, Onoe Y, et~al (2007) {Parameter Estimation of PEMFC by
  On-Line Tracking Simulator}. In: {SICE Annual Conference 2007}, IEEE, pp
  2946--2949

\bibitem[{Nakaya et~al(2008)Nakaya, Seki, Kawaguchi, Onoe, and
  Ootani}]{nakaya2008}
Nakaya M, Seki T, Kawaguchi K, et~al (2008) {Model parameter estimation by
  tracking simulator for the innovation of plant operation}. IFAC Proceedings
  Volumes 41(2):2168--2173

\bibitem[{Nakaya et~al(2009)Nakaya, Nakabayashi, Ohtani, Ikegaya, Asawa,
  Terashima, Izawa, and Ogata}]{nakaya2009}
Nakaya M, Nakabayashi A, Ohtani T, et~al (2009) {A new estimation method by
  utilizing on-line tracking simulator}. In: {2009 ICCAS-SICE}, IEEE, pp
  3274--3277

\bibitem[{Nakaya et~al(2011)Nakaya, Ikegaya, Nakabayashi, Ootani, Chen, Li,
  Wang, and Ogata}]{nakaya2011}
Nakaya M, Ikegaya Y, Nakabayashi A, et~al (2011) {Online process simulator with
  hybrid model of physical model and just-in-time model}. IFAC Proceedings
  Volumes 44(1):1640--1645

\bibitem[{Pantelides and Renfro(2013)}]{pantelides2013}
Pantelides CC, Renfro JG (2013) {The online use of first-principles models in
  process operations: Review, current status and future needs}. Computers \&
  Chemical Engineering 51:136--148

\bibitem[{Papaioannou et~al(2016)Papaioannou, Sutton, and
  Booth}]{papaioannou2016}
Papaioannou D, Sutton A, Booth A (2016) {Systematic approaches to a successful
  literature review}. Systematic approaches to a successful literature review
  pp 1--336

\bibitem[{Petschnigg et~al(2018)Petschnigg, Breitenhuber, Breiling, Dieber, and
  Brandst{\"o}tter}]{petschnigg2018}
Petschnigg C, Breitenhuber G, Breiling B, et~al (2018) {Online simulation for
  flexible robotic manufacturing}. In: {Int. Conf. Ind. Technol. Manag}, pp
  88--92

\bibitem[{Pietil{\"a} et~al(2013)Pietil{\"a}, Kaartinen, and
  Reinsalo}]{pietila2013}
Pietil{\"a} J, Kaartinen J, Reinsalo AM (2013) {Parameter estimation for a
  flotation process tracking simulator}. IFAC Proceedings Volumes
  46(16):122--127

\bibitem[{Pujo et~al(2004)Pujo, Pedetti, and Ounnar}]{pujo2004}
Pujo P, Pedetti M, Ounnar F (2004) {Pilotage proactif des lignes de production
  kanban par modelisation DEVS et simulation temps reel}. In: 5e Conference
  Francophone de MOdelisation et SIMulation - Modelisation et simulation pour
  l'analyse et l'optimisation des systemes industriels et logistiqes, MOSIM'04,
  Nantes, France

\bibitem[{Rosen et~al(2020)Rosen, J{\"a}kel, Barth, Stern, Schmidt-Vollus,
  Heinzerling, Hoffmann, Richter, Schmidt, Scheifele et~al}]{rosen2020}
Rosen R, J{\"a}kel J, Barth M, et~al (2020) {Simulation und digitaler Zwilling
  im Anlagenlebenszyklus}. VDI Statusreport 1

\bibitem[{Ruusu et~al(2017)Ruusu, Santill{\'a}n~Mart{\'\i}nez, Karhela, and
  Vyatkin}]{ruusu2017}
Ruusu R, Santill{\'a}n~Mart{\'\i}nez G, Karhela T, et~al (2017) {Sliding mode
  SISO control of model parameters for implicit dynamic feedback estimation of
  industrial tracking simulation systems}. In: {IECON 2017-43rd Annual
  Conference of the IEEE Industrial Electronics Society}, IEEE, pp 6927--6932

\bibitem[{Santill{\'a}n~Mart{\'\i}nez et~al(2015)Santill{\'a}n~Mart{\'\i}nez,
  Karhela, Niemist{\"o}, Rossi, Pang, and Vyatkin}]{martinez2015}
Santill{\'a}n~Mart{\'\i}nez G, Karhela T, Niemist{\"o} H, et~al (2015) {A
  hybrid approach for the initialization of tracking simulation systems}. In:
  {2015 IEEE 20th Conference on Emerging Technologies \& Factory Automation
  (ETFA)}, IEEE, pp 1--8

\bibitem[{Santill{\'a}n~Mart{\'\i}nez et~al(2017)Santill{\'a}n~Mart{\'\i}nez,
  Karhela, Ruusu, Lackman, and Vyatkin}]{martinez2017}
Santill{\'a}n~Mart{\'\i}nez G, Karhela T, Ruusu R, et~al (2017) {Towards a
  systematic path for dynamic simulation to plant operation: OPC UA-enabled
  model adaptation method for tracking simulation}. In: {IECON 2017-43rd Annual
  Conference of the IEEE Industrial Electronics Society}, IEEE, pp 5503--5508

\bibitem[{Santill{\'a}n~Mart{\'\i}nez
  et~al(2018{\natexlab{a}})Santill{\'a}n~Mart{\'\i}nez, Karhela, Ruusu, Sierla,
  and Vyatkin}]{martinez2018b}
Santill{\'a}n~Mart{\'\i}nez G, Karhela TA, Ruusu RJ, et~al (2018{\natexlab{a}})
  {An integrated implementation methodology of a lifecycle-wide tracking
  simulation architecture}. IEEE Access 6:15391--15407

\bibitem[{Santill{\'a}n~Mart{\'\i}nez
  et~al(2018{\natexlab{b}})Santill{\'a}n~Mart{\'\i}nez, Sierla, Karhela, and
  Vyatkin}]{martinez2018a}
Santill{\'a}n~Mart{\'\i}nez G, Sierla S, Karhela T, et~al (2018{\natexlab{b}})
  {Automatic generation of a simulation-based digital twin of an industrial
  process plant}. In: {IECON 2018-44th Annual Conference of the IEEE Industrial
  Electronics Society}, IEEE, pp 3084--3089

\bibitem[{Saptoro(2014)}]{saptoro2014}
Saptoro A (2014) {State of the art in the development of adaptive soft sensors
  based on just-in-time models}. Procedia Chemistry 9:226--234

\bibitem[{Scheer et~al(2021)Scheer, Straßburger, and Knapp}]{scheer2021}
Scheer R, Straßburger S, Knapp M (2021) {Digital-physische Verbundkonzepte:
  Gegen-{\"u}berstellung, Nutzeffekte und kritische H{\"u}rden}. Cuvillier
  Verlag, p~11

\bibitem[{Schumann et~al(2013)Schumann, Witt, and Klimant}]{schumann2013}
Schumann M, Witt M, Klimant P (2013) {A real-time collision prevention system
  for machine tools}. Procedia CIRP 7:329--334

\bibitem[{Seki et~al(2008)Seki, Fukano, Kawaguchi, Nakabayashi, Hatsugai,
  Nakaya, Ohtani, Yokoyama, Kawamura, and Oguchi}]{seki2008}
Seki T, Fukano G, Kawaguchi K, et~al (2008) {Innovative plant operations by
  using tracking simulator}. In: {2008 SICE Annual Conference}, IEEE, pp
  2100--2103

\bibitem[{Sekler and Verl(2009)}]{sekler2009}
Sekler P, Verl A (2009) {Real-time computation of the system behaviour of
  lightweight machines}. In: {2009 First International Conference on Advances
  in System Simulation}, IEEE, pp 144--147

\bibitem[{Sekler et~al(2012)Sekler, Vo{\ss}, and Verl}]{sekler2012}
Sekler P, Vo{\ss} M, Verl A (2012) {Model-based calculation of the system
  behavior of machine structures on the control device for vibration
  avoidance}. The International Journal of Advanced Manufacturing Technology
  58(9-12):1087--1095

\bibitem[{Svensson et~al(2012)Svensson, Danielsson, and
  Lennartson}]{svensson2012}
Svensson B, Danielsson F, Lennartson B (2012) {Time-synchronised
  hardware-in-the-loop simulation—Applied to sheet-metal press line
  optimisation}. Control Engineering Practice 20(8):792--804

\bibitem[{Tjahjono et~al(2013)Tjahjono, Teixeira, and Alfaro}]{tjahjono2013}
Tjahjono B, Teixeira ELS, Alfaro SCA (2013) {An online simulation to link asset
  condition monitoring and operations decisions in through-life engineering
  services}. In: {2013 Winter Simulations Conference (WSC)}, IEEE, pp 159--168

\bibitem[{Webster and Watson(2002)}]{webster2002}
Webster J, Watson RT (2002) {Analyzing the past to prepare for the future:
  Writing a literature review}. MIS quarterly pp xiii--xxiii

\bibitem[{Yoshitani et~al(1991)Yoshitani, Naganuma, and Yanai}]{yoshitani1991}
Yoshitani N, Naganuma Y, Yanai T (1991) {Optimal slab heating control for
  reheating furnaces}. In: {1991 American Control Conference}, IEEE, pp
  3030--3035

\bibitem[{Zipper(2021{\natexlab{a}})}]{zipper2021a}
Zipper H (2021{\natexlab{a}}) {Method for synchronisation of
  online-simulation}. at-Automatisierungstechnik 69(11):1020--1021

\bibitem[{Zipper(2021{\natexlab{b}})}]{zipper2021b}
Zipper H (2021{\natexlab{b}}) {Real-time-capable synchronization of Digital
  Twins}. IFAC-PapersOnLine 54(4):147--152

\bibitem[{Zipper and Diedrich(2019)}]{zipper2019}
Zipper H, Diedrich C (2019) {Synchronization of industrial plant and digital
  twin}. In: {2019 24th IEEE international conference on emerging technologies
  and factory automation (ETFA)}, IEEE, pp 1678--1681

\bibitem[{Zipper et~al(2018)Zipper, Auris, Strahilov, and Paul}]{zipper2018}
Zipper H, Auris F, Strahilov A, et~al (2018) {Keeping the digital twin
  up-to-date—Process monitoring to identify changes in a plant}. In: {2018
  IEEE International Conference on Industrial Technology (ICIT)}, IEEE, pp
  1592--1597

\bibitem[{Zupan et~al(2021)Zupan, {\v{S}}imic, and Herakovi{\v{c}}}]{zupan2021}
Zupan H, {\v{S}}imic M, Herakovi{\v{c}} N (2021) {Realization of an Optimal
  Production Plan in a Smart Factory with On-line Simulation}. In: {Service
  Oriented, Holonic and Multi-Agent Manufacturing Systems for Industry of the
  Future: Proceedings of SOHOMA 2020}, Springer, pp 485--495

\end{thebibliography}
